# Sensitive and accurate dual wavelength UV-VIS polarization detector for optical remote sensing of tropospheric aerosols

G. David, A. Miffre, B. Thomas and P. Rairoux

**Grégory David**[1], (gdavid@lasim.univ-lyon1.fr)

**Alain Miffre**[1*], (miffre@lasim.univ-lyon1.fr)

**Benjamin Thomas**[1], (bthomas@lasim.univ-lyon1.fr)

**Patrick Rairoux**[1], (rairoux@lasim.univ-lyon1.fr)

[1] *Laboratoire de Spectrométrie Ionique et Moléculaire, CNRS, UMR 5579 Université Lyon 1,*

*10 rue da Byron, 69622 Villeurbanne, France*

* Corresponding author:

miffre@lasim.univ-lyon1.fr

fon: 0033-472.43.10.87

fax: 0033-472-43.15.07

**Abstract**

An UV-VIS polarization Lidar has been designed and specified for aerosols monitoring in the troposphere, showing the ability to precisely address low particle depolarization ratios, in the range of a few percents. Non-spherical particle backscattering coefficients as low as $5 \times 10^{-8}$ $m^{-1}.sr^{-1}$ have been measured and the particle depolarization ratio detection limit is 0.6 %. This achievement is based on a well-designed detector with laser-specified optical components (polarizers, dichroic beamsplitters) summarized in a synthetic detector transfer matrix. Hence, systematic biases are drastically minimized. The detector matrix being diagonal, robust polarization calibration has been achieved under real atmospheric conditions. This UV-VIS polarization detector measures particle depolarization ratios over two orders of magnitude, from 0.6 up to 40 %, which is new, especially in the UV where molecular scattering is strong. Hence, a calibrated UV polarization-resolved time-altitude map is proposed for urban and free tropospheric aerosols up to 4 kilometres altitude, which is also new. These sensitive and accurate UV-VIS polarization-resolved measurements enhance the spatial and time evolution of non-spherical tropospheric particles, even in urban polluted areas. This study shows the capability of polarization-resolved laser UV-VIS spectroscopy to specifically address the light backscattering by spherical and non-spherical tropospheric aerosols.

**Keywords**

Polarization, UV-VIS Lidar, aerosols, Non-spherical particles.

## 1. Introduction

Atmospheric aerosols (or suspended particulate matter, PM) play a key role in the Earth atmosphere radiative balance both directly, by light extinction, and indirectly, through complex processes involving aerosols physical and chemical properties [1]. A well-known example is given in the stratosphere by the ozone layer destruction in Polar Regions, related to anthropogenic polar stratospheric clouds through complex photo-catalytic surface reactions [2-4]. In the troposphere, atmospheric particles affect hydrometeor formation by acting as condensation nuclei [5]. Recent studies have shown that PM may also modify the physical and chemical properties of the atmospheric Planetary Boundary Layer (PBL), especially in urban polluted areas [6]. For climate forcing assessments, the indirect role of particles on the radiative properties of atmospheric particles must be quantitatively estimated. This task is very difficult since a complete physical and chemical PM characterization is not yet achieved [7]. In particular, there is a need for measuring spatial and temporal variations of PM-concentrations in urban polluted areas. One of the major limitations originates from the lack of detailed knowledge on atmospheric aerosols optical properties, which induces large uncertainties.

To face such a complexity, laser spectroscopy is of prime importance since PM light scattering and extinction are nowadays the main applied optical properties to evaluate the PM-atmospheric content. Several optical properties can be addressed by laser spectroscopy, as for example the scattering phase function [8]. Laser spectroscopy can be applied to address the atmospheric particles chemical composition, by studying their refractive index wavelength dependency [9]. Laser induced fluorescence is another methodology to access to PM chemical composition also used to characterize organic and biogenic atmospheric particles [10]. Along with these laser techniques, laser active remote sensing (Lidar) is particularly interesting as it provides fast, reliable and range-resolved access to the optical properties of an ensemble of atmospheric particles, under atmospheric conditions of temperature and humidity [11, 12]. The laser excitation wavelength λ is often chosen in the visible (VIS) or/and in the infra-red (IR) spectral range [10, 12, 13] while the ultraviolet (UV) spectral range is rather seldom used [14-16]. In the meantime, atmospheric particles present a three-modal size distribution with an ultra-fine mode (in the nanometer scale), a fine mode (in the range of a few cents of nanometers) and a coarse mode (in the micrometer size range). While coarse particles experience sedimentation processes and ultra-fine particles encounter aggregation and condensation processes, fine particles have the longest lifetime in the atmosphere (several days). Hence, most observed atmospheric particles are fine and in urban polluted areas, the number of fine particles can exceed 1 000 part.$cm^{-3}$ [6]. To address high concentrated fine particles with laser remote sensing, it is interesting to choose a laser excitation wavelength in the UV spectral range, where particles size parameters ($x = 2\pi r/\lambda$ for an r equivalent sphere radius) often lead to scattering phase function enhancements [17]. This is however challenging since in the UV spectral range, molecular scattering may overcome particles scattering.

Among the major uncertainties involved in climate change modeling, the lack of knowledge on the atmospheric particles shape is an essential point, especially in urban polluted areas, where atmospheric aerosols may present a wide range of sizes and shapes. Applying the century-old Lorenz-Mie formalism to tropospheric particles may lead to significant errors in climate change modeling [18], as non-spherical particles scatter light differently from volume or surface-equivalent spheres. In particular, orientation averaging over an ensemble of non-spherical particles does not lead to the same scattering pattern as for spheres [19]. Hence, non-spherical particles are difficult to address since no general analytical solution is available, except for some specific geometry far away from the observed highly-irregularly shape of atmospheric particles [17]. However, in the Lidar backward direction, the polarization of the phase function is unequivocally sensitive to particles shape modifications [17], which makes the detection of this property attractive for laser remote sensing. For spherical particles, the polarization state of the laser is preserved during the scattering process. In contrast, scattering of light by an ensemble of randomly-oriented non-spherical particles modify the polarization state of the laser. More precisely, as this polarization state is analyzed at a far range compared to the particles size, randomly-oriented non-spherical particles exhibit a non-zero polarization change [17], often called depolarization, whose magnitude (hereafter called the particles depolarization ratio) is a signature of the particles shape. Hence, polarization-sensitive Lidar systems can be used as particles shape indicators.

In this paper, we remotely address the polarization optical properties of tropospheric particles with a new home-built dual-wavelength (UV-VIS) polarization Lidar. The addressed particles are present in the PBL and in the free troposphere. The UV-light has been chosen to improve our sensitivity to the fine particles mode while the dual-wavelength (UV-VIS) enables to address the spectral properties of tropospheric urban aerosols. To our knowledge, tropospheric urban aerosols have only been studied in the VIS or / and the IR spectral range [13,20]. Recent studies mainly focused on the remote measurement of high particles depolarization ratios, in the range of 40 % as observed during volcanic ash intrusion episodes [21-23] or in the 20 %-range, as for Saharan desert dust particles intrusion events [24]. In between these rather seldom episodes, the particles load of an urban troposphere is usually dominated by local sources of particulate matter, originating from petrochemical plants and traffic-jam polluting the urban canopy [25]. There is no a priori evidence that such atmospheric PBL particles be spherical. Rather low depolarization ratios, in the range of a few percents, may be expected for urban tropospheric particles, but the magnitude of this depolarization needs to be accurately evaluated to detect very small changes in the particles optical properties in general, and in the particles shape in particular. Hence, in this paper, we concentrate on the measurement of low particles depolarization ratios, in the range of few percents, which is the most frequently observed situation in the Lyon troposphere. To trustworthy measure such low depolarization ratios, the polarization detector must be very sensitive and very accurately designed. This is challenging since a small system bias in the measured depolarization ratio, originating for example from an imperfection on an optical component, may lead to substantial errors in the measured depolarization ratio. Accordingly, the Lidar laser source and the polarization detector must be very accurately specified. In this paper, a systematic study is proposed to specify the spectral and the polarization properties of each optical component used in the dual-wavelength polarization detector. This specification has been performed on a laboratory dedicated test bench. In the literature, manufacturer's specifications are often trusted. Here, our approach is different since we combine UV and VIS-polarization measurement in a single detector while optical specifications are rarely at the same level in these spectral ranges and often given for continuous unpolarized white light, instead of the monochromatic polarized pulses used in a polarization Lidar. Moreover, we are interested in (UV-VIS) low depolarization ratios measurements so that any system bias must be carefully analyzed.

The novelty of this work is hence threefold. First, we consider the UV-VIS Lidar polarization observation with depolarization ratios in the percent range, which is a very low value for atmospheric Lidar observation. It induces strong constraints on the dual-wavelength polarization Lidar experimental set-up and on its calibration. Secondly, the biases in the depolarization ratio measurements are quantitatively estimated by considering the state-of-the-art for optical components in the UV spectral range. Within our home-built polarization detector, a detection limit of $6 \times 10^{-3}$, comparable to the molecular depolarization, is achievable for remote polarization measurements. Thirdly, examples of sensitive and accurate Lidar depolarization are presented in the PBL and the free troposphere and analyzed in terms of PM laser light scattering in the UV and VIS-spectral range. The paper is organized as follows. Section 2 is dedicated to theoretical considerations. Starting from scattering of light by atmospheric particles, we analyze several possible system biases affecting the depolarization ratio measurement at UV-VIS wavelengths. The sensitivity of these biases for measuring low depolarization ratios is analyzed and for the first time, the role of the dichroic beamsplitter, used for dual-wavelength polarization measurements, is addressed. To optimize our sensitivity to low particle depolarization ratios, the sky background contribution to the Lidar signal is then analyzed and polarization-resolved. In section 3, we first present our Lidar experimental set-up (emitter, receiver and detector). Then, to fulfill the requirements derived in section 2, the spectral and polarization properties of each detector optical component are specified in our laboratory. A detector transfer matrix is provided to underline the performances of our dual-wavelength polarization detector. As an output, we present in section 4 experimental measurements of tropospheric particles depolarization in the percent range, in the UV and in the VIS-spectral range. To our knowledge for the first time, a calibrated UV polarization-resolved time-altitude map is proposed for tropospheric urban aerosols. The paper ends with a conclusion and outlooks.

## 2. Dual-wavelength polarization Lidar methodology

In this section, we focus on theoretical considerations for retrieving the particle backscattering coefficient $\beta_p$ and the particle depolarization ratio $\delta_p$. In particular, we quantitatively analyze the possible systematic biases affecting the $\delta_p$-measurement at two wavelengths, by focusing on $\delta_p$-measurements in the range of a few percents.

### 2.1 Scattering of light by atmospheric particles

Light-scattering by an ensemble of particles, either spherical or not, can be described in the frame of the scattering matrix **F**, which relates the Stokes parameters of the incident and detected scattered light beams [17]. In this formalism, an incident light wave (wave-vector $\mathbf{k_i}$, polarization $\boldsymbol{\pi}_\mathbf{i}$) is scattered by an ensemble of particles of arbitrary size and shape in all directions, the detector direction defining the scattering angle θ and the scattering plane between the incident and scattered light waves (wave-vector **k**, polarization **π**). The $\boldsymbol{\pi}_i$ and **π**-polarization states are usually defined with respect to the light scattering plane, either parallel (p) or perpendicular (s) to this plane. In this paper, our main concern is on the polarization state of the light wave backscattered (θ = 180°) by tropospheric particles, probed with a linearly polarized laser beam propagating through the atmosphere. In this Lidar backscattering case (θ = 180°), for randomly oriented particles, the **F**-matrix is diagonal and only depends on its first two elements $F_{11}$ and $F_{22}$, which results in a linear depolarization ratio δ :

$$\delta = (F_{11} - F_{22}) / (F_{11} + F_{22}) \quad (1)$$

The depolarization ratio δ, determined by the $F_{22}/F_{11}$-ratio, is hence an intrinsic property of randomly-oriented particulate matter, mainly governed by the particles shape [17]. It may also depends on the laser wavelength λ and, as for $F_{11}$ and $F_{22}$, on the size parameter x and on the complex refractive index m [24]. Spherical particles, for which $F_{11} = F_{22}$, induce no depolarization (δ = 0) in contrary to non-spherical particles for which the equality $F_{11} = F_{22}$ no longer holds. Hence, the δ-ratio can be used as a non-sphericity indicator of an ensemble of particles [17].

### a. Backscattering and depolarization in the atmosphere

At altitude z above ground, the polarization components of the wave backscattered by the atmosphere (intensity $\mathbf{I} = [I_p, I_s]^T$) are related to the polarization components of the incident laser wave (intensity $\mathbf{I}_i = [I_{p,i}, I_{s,i}]^T$) through the well-known Lidar equation [26,27], assuming single-scattering from the atmosphere :

$$\boldsymbol{I}(\lambda, z) = \frac{T^2(z)}{z^2}[\boldsymbol{\beta}](\lambda, z)\boldsymbol{I}_{\mathrm{i}} + \mathbf{I}_{\mathrm{sb}} \quad \text{with} \quad \beta(\lambda, z) = \begin{bmatrix} \beta_{//} & \beta_{\perp} \\ \beta_{\perp} & \beta_{//} \end{bmatrix} \tag{2}$$

for the specific case of Lidar elastic backscattering. The intensity is here considered instead of power or photons numbers to be independent of the surface detector. The Lidar equation is presented in the form of column-vectors to facilitate the discussion proposed in section 2.2. In the Lidar backscattering case, the incident laser linear polarization is often taken as a reference so that p or s-polarized components are preferably referred to as parallel (//) or perpendicular (⊥) with respect to the laser linear polarization. Hence, the **β**-matrix coefficients are defined with respect to the laser linear polarization. As shown in section 3, the relationship between the backscattered intensity vector **I** and the detected intensity vector $\mathbf{I}^* = [I_{//}, I_{\perp}]^T$ can be expressed as follows:

$$I^*(\lambda,z) = O(z) \times [\boldsymbol{\eta}(\boldsymbol{\lambda})]\ \mathbf{I}(\lambda,z) \tag{3}$$

where O(z) is the overlap function to be specified in section 3.1 and $[\boldsymbol{\eta}(\boldsymbol{\lambda})]$ is the detector transfer matrix corresponding to the excitation laser wavelength λ, to be specified in section 3.3. Finally, T(λ,z) denotes the optical transmission of the atmosphere and the intensity vector $\mathbf{I}_{sb} = [I_{sb,p}, I_{sb,s}]^T$ represents the sky background contribution to the intensity Lidar signal **I**, as described in section 2.3.

The Lidar signal **I** results from atmospheric molecules $N_2$ and $O_2$ (subscript m) and particles (subscript p) backscattering. Application of the superposition principle to the volume backscattering coefficient β implies that $\beta = \beta_m + \beta_p$. As shown in [24], the particles backscattering coefficient β is linked to the scattering matrix **F** by $\beta_{//} = (F_{11} + F_{22})/2$ and $\beta_{\perp} = (F_{11} - F_{22})/2$, so that, following equation (1), the atmosphere depolarization ratio δ is usually defined as: $\delta = \beta_{\perp}/\beta_{//}$. While the β-backscattering coefficient is additive (as for $F_{11}$ and $F_{22}$), the δ-ratio is an intensive parameter.

However, in the atmosphere, both molecules and particles, which are a priori non-spherical, contribute to the depolarization. Hence, a molecular ($\delta_m$) and a particle ($\delta_p$) depolarization ratio can be defined. The molecular depolarization is due to molecules anisotropy, which provokes the apparition of Raman ro-vibrational sidebands in the molecular backscattering spectrum, responsible for light depolarization [28,29]. The relationship between $\delta$, $\delta_p$ and $\delta_m$ has been first proposed in [30]:

$$\delta = \left(1 - \frac{1}{R_{//}}\right)\delta_p + \frac{\delta_m}{R_{//}} \quad (4)$$

where $R_{//} = 1 + \beta_{p,//}/\beta_{m,//}$ is known as the parallel Lidar R-ratio, representing the contrast of molecular backscattering compared to particles backscattering (a particle-free or molecular atmosphere satisfies to $R_{//} = 1$). As a result of the well-known $\lambda^{-4}$ Rayleigh law, molecular backscattering in the UV-spectral range (at $\lambda = 1064/3$ nm) is approximately five times more intense than in the VIS-spectral range (at $\lambda = 1064/2$ nm). Hence, particles backscattering $\beta_p$ and particles depolarization ratios $\delta_p$ are rather difficult to measure in the UV and a higher sensitivity and accuracy for the ($\beta_p$, $\delta_p$)-measurement are needed. Hence, in the literature, $\delta$ is sometimes preferably measured rather than $\delta_p$ [15]. In a molecular atmosphere, vertical profiles of $\beta_{m,//}$ and $\beta_{m,\perp}$ have been determined from molecular scattering computation, using reanalysis model from the European Centre for Medium-Range Weather Forecasts (ECWMF). As shown by A. Behrendt [29], $\beta_m$ and $\delta_m$ depend on the detector daylight filter bandwidth ($\Delta\lambda$). Thanks to the spectral selectivity of our detector, $\delta_m$ has negligible temperature dependence and deduced from molecular scattering theory. For $\Delta\lambda = 0.35$ nm, we get $\delta_m = 3.7 \times 10^{-3}$ at $\lambda = 355$ nm and $3.6 \times 10^{-3}$ at $\lambda = 532$ nm.

**b. Particle backscattering and depolarization ratio ($\beta_p$, $\delta_p$)-retrieval methodology**

In this paragraph, the methodology to derive $\beta_p$ and $\delta_p$ is described. As shown in equation (4), $R_{//}$ and $\delta$ have to be measured. The parallel Lidar R-ratio is computed by applying the Klett's inversion algorithm [31] to correct for the particles extinction in the Lidar equation. A predefined value for the particles backscatter-to-extinction ratio $S_p$ is needed as well as a starting point $z_0$ for the inversion algorithm, generally chosen at high altitudes. As detailed in [32], $S_p$ depends on the particles microphysics and is a priori varying with z-altitude. In the free troposphere, $S_p$-values of 50 sr have been reported in the literature [33] and chosen in our inversion algorithm with an error bar of 5 sr. In the Planetary Boundary Layer (PBL), moisture effects and chemical composition may strongly influence the particles size distribution and consequently the $S_p$-value. We numerically calculated $S_p$ as a function of the relative humidity by using Mie and Rayleigh-Gans theory, for a realistic three-mode particles size distribution detailed in [25], including soot, organics, sulphate and silicate particles. In between the PBL and the free troposphere, we assumed a linear variation of $S_p$ with z-altitude starting at observed inversion layers. At altitude z, the accuracy on the Lidar ratio is derived from the Klett's algorithm, by using the maximum and minimum values of $S_p$.

Vertical profiles of $\beta_{p,//}$ are retrieved from the parallel Lidar R-ratio computation by applying the Klett's algorithm to the parallel Lidar intensity signal $I_{//}$. The $\beta_{p,\perp}$–coefficient is very interesting to derive as it is non-spherical particles specific, in contrary to $\beta_{p,//}$ and $\delta_p$. By using equation (4) and the $\delta_p$-definition ($\delta_p = \beta_{p,\perp}/\beta_{p,//}$), we derive $\beta_{p,\perp} = (R_{//}\delta - \delta_m) \times \beta_{m,//}$, providing vertical profiles of $\beta_{p,\perp}$ as a function of z-altitude. The uncertainty on $\delta_p$ is derived from equation (4) and expresses as follows:

$$\frac{\Delta\delta_p}{\delta_p} = \frac{R_{//}\Delta\delta}{R_{//}\delta - \delta_m} + \frac{\delta\Delta R_{//}}{R_{//}\delta - \delta_m} + \frac{\Delta R_{//}}{R_{//} - 1} \qquad (5)$$

## 2.2 Theoretical considerations for remote sensing of low depolarization ratios

The emission and the receiver Lidar systems, which are polarization-sensitive, may modify our perception of the polarization backscattered by the atmosphere. To unambiguously determine depolarization ratios, we have to correct for this system biases affecting the depolarization ratio measurement at two wavelengths. Literature in this field is quite abundant. Pioneer work has been done by J. Biele et al. [34] who developed an algorithm to remove the effect of a cross-talk component small compared to the observed depolarization, in the case of polar stratospheric clouds (PSC). Then, Adachi et al. [26] used a calibration method for accurate estimation of PSC depolarization ratios estimates. In 2003, J. Reichardt et al. developed a method for determining δ by using three elastic-backscatter Lidar signals [35]. In 2006, J.M. Alvarez developed a three-measurement method to calibrate polarization-sensitive Lidars [36], further extended by V.F. Freudenthaler et al. in 2009 to the case of desert dust particles [16]. Finally, the specific case of a single-channel detector used to measure both polarization components has been studied in [37]. Each of these algorithm correction schemes is well-suited for its designed case. Here, our main concern is dedicated to low δ-measurements, in the range of a few percents. It is the subject of this paragraph to quantitatively evaluate the system constraints to measure these low depolarization ratios. An atmosphere having a low depolarization ratio δ, in the range of a few percents, is hence considered as an input. In the absence of undesirable system bias, the measured depolarization δ* would be equal to δ but in general, δ* differs from δ. Relations between δ* and δ are here provided to account for several sources of systematic errors presented in figure 1. The role of the dichroic beamsplitter, introduced for dual-wavelength detection is analyzed in details. Each system bias is studied separately to specifically address its contribution to δ*, hence quantifying the relative error between δ* and δ, for δ-values in the percent range.

**Please insert figure 1 here.**

**a. Influence of a small unpolarized polarization component emitted in the atmosphere**

Here, we quantify the effect of a small unpolarized component in the emitted laser polarization on the δ-measurement. This emitted unpolarized component may originate from the laser polarization purity or / and from polarization-sensitive reflective mirrors from the emission optics. When the polarization state of the emitted laser has two polarization components (i.e. $I_i = I_{i,//} + I_{i,\perp}$) , a residual polarization $\varepsilon = I_{i,\perp}/ I_i \ll 1$ is emitted throughout the atmosphere (see figure 1a). In this case, even in a non-depolarizing atmosphere ($\delta = 0$), the polarization state of the backscattered wave will have a depolarized component, leading to a non-zero measured depolarization $\delta^*$, i.e. $\delta^* \geq \varepsilon$. Equation (2) shows that the parallel Lidar intensity $I_{//}$ is contaminated by the induced non-zero $\beta_{\perp} I_{i,\perp}$ term while the perpendicular Lidar intensity $I_{\perp}$ is contaminated by the term $\beta_{\perp} I_{i,//}$. Hence, after a few calculations, $\delta^*$ can be expressed as a function of δ and ε, the bias parameter, as follows:

$$\delta^* = \frac{(1-\varepsilon)\,\delta + \varepsilon}{(1-\varepsilon) + \varepsilon\delta} \qquad (6)$$

When $\delta = 10$ %, a residual polarization $\varepsilon = 1$ % induces a measured depolarization $\delta^* = 11$ %. Moreover, as shown by equation (6) plotted in figure 1a, care should be taken when measuring low depolarization ratios, in the 1 %-range: for $\delta = 1$ %, the required ε-value to ensure that $\delta^*$ differs from δ by no more than 1 %, is only equal to $10^{-2}$ %.

**b. Imperfect separation of polarization components, polarization cross-talks**

When separating the two polarization components $\boldsymbol{\pi} = \{//,\perp\}$, defined with respect to the laser linear polarization, some leakage between the two polarization detection channels may occur, leading to an imperfect polarization separation through cross-talk effects. To calculate the allowed leakage for measuring δ-values in the range of a few percents, we introduce a cross-talk coefficient $CT_{//}$ to characterize the leakage of the //-polarization channel into the ⊥-polarization channel. As shown in figure 1b, the parallel Lidar intensity $I_{//}$ is contaminated by the contribution from perpendicular channel, having a $CT_{\perp}$-efficient, while removing the leakage contribution into the perpendicular channel, which occurs with a $CT_{//}$-efficiency. Hence, the measured parallel Lidar intensity $I^*_{//}$ is given by: $I^*_{//} = (1 - CT_{//})I_{//} + CT_{\perp}I_{\perp}$. Symmetrically, the perpendicular Lidar intensity can be written as $I^*_{\perp} = (1 - CT_{\perp})\, I_{\perp} + CT_{//}\, I_{//}$, as obtained from the $I^*_{//}$-expression by simply exchanging the // and ⊥-subscripts, to satisfy photon energy conservation, hence introducing the $CT_{\perp}$ cross-talk coefficient, characterizing the leakage of the ⊥-polarization channel into the //-polarization channel. δ* is linked to δ via the bias parameters $CT_{//}$ and $CT_{\perp}$ as follows:

$$\delta^* = \frac{(1-CT_{\perp})\,\delta + CT_{//}}{(1-CT_{//}) + \delta\ CT_{\perp}} \qquad (7)$$

For a δ = 10 % atmospheric input depolarization, a bias parameter of $CT_{//} = CT_{\perp} = 1$ % leads to δ* = 11 %. As shown by equation (7) and in figure 1b, care should be taken when measuring low depolarization ratios, in the 1 %-range: for δ = 1 %, the same residual leakage induces a measured depolarization δ* of 2 %, which represents a 100 %-relative error.

**c. Misalignment between the transmitter and receiver polarization axes**

The polarization of backscattered photons is analyzed by projection on the polarization axes of the Lidar detector. It is implicitly assumed that these polarization axes merge with the laser linear polarization axes, so that the polarization plane of the transmitter and the receiver are in perfect alignment. When a systematic offset-angle φ exists between the emitter and receiver axes (see figure 1c), as first described by J.M. Alvarez [36], the measured depolarization δ* can be expressed as a function of δ and the φ-angle as follows:

$$\delta^* = G_\lambda \frac{\delta + \tan^2(\varphi)}{1+\delta\tan^2(\varphi)} \quad (8)$$

where $G_\lambda$ is the electro-optics calibration constant to be specified in section 3.4. The relative error bar on δ is plotted in figure 1c for different φ angles. When δ = 10 %, a residual offset angle of 5° leads to δ* = 10.7 % only. For δ = 1 %, when φ = 1° (resp. 5°), δ* = 1.03 % (resp. 1.76 %). Varying the offset angle φ can be used to calibrate our depolarization measurements to determine $G_\lambda$, as proposed by J.M. Alvarez [36] and as detailed in section 3.4.

**d. Possible influence of a dichroic beamsplitter**

In dual-wavelength polarization Lidar detectors, a dichroic beamsplitter is often introduced to differentiate the polarization state of backscattered photons at the two laser wavelengths. In this paragraph, we analyze the possible bias introduced by such a dichroic beamsplitter on the measurement of low depolarization ratios δ. To our knowledge, such a systematic study has never been reported in the literature, where the dichroic beamsplitter is assumed to be polarization-insensitive.

Let us consider a dichroic beamsplitter having $R_p$, $R_s$-reflectivity coefficients, defined with respect to the dichroic beamsplitter incidence plane (a similar discussion could be held on the corresponding transmission coefficients ($T_p = 1 - R_p$, $T_s = 1 - R_s$) coefficients). As a consequence of Fresnel's formula, $R_p$ generally differs from $R_s$ ($R_p < R_s$), so that the reflection on the dichroic beamsplitter does not modify the polarization state of the backscattered photons which remains linear but is rotated. However, in the absence of polarization cross-talks, the dichroic beamsplitter induces no leakage between the two polarization channels so that the difference in $R_p$, $R_s$-values is generally simply taken into account during the polarization calibration procedure as a multiplicative constant.

The above situation implicitly assumes that the laser linear polarization axes merge with the p and s-axes of the dichroic beamsplitter, so that the polarization plane of the transmitter and the dichroic beamsplitter are in perfect alignment. When a systematic offset-angle $\theta_0$ exists between the parallel laser linear polarization axis and the p-axis of the dichroic beamsplitter (see figure 1d), polarization cross-talks appear, which cannot be compensated during the polarization calibration procedure. We quantified the effect of a non-zero offset angle $\theta_0$ on the measurement of a low atmosphere depolarization ratio $\delta$. The corresponding calculations are detailed in appendix. In the presence of a non-zero offset angle $\theta_0$, $\delta^*$ differs from $\delta$:

$$\delta^* = \frac{a^2 cos^2\theta_0 sin^2\theta_0 + \delta_0 (b - a cos^2\theta_0)^2}{(b - a sin^2\theta_0)^2 + \delta_0\, a^2 cos^2\theta_0 sin^2\theta_0} \quad (9)$$

where the two coefficients $a = \sqrt{R_p} - \sqrt{R_s}$ and $b = \sqrt{R_p}$ are determined by the dichroic beamsplitter $R_p$, $R_s$-reflectivity coefficients, as detailed in the appendix. When there is no offset angle, $\delta^*$ is proportional to $\delta$ so that the corresponding proportionality coefficient $R_s/R_p$ can be included in the polarization calibration procedure. The relative error bar on $\delta$ is plotted in figure 1d for different offset angles $\theta_0$, using $R_p$ = 72 % and $R_s$ = 94 %.

### 2.3 Sky background contribution to the polarized Lidar signal

Scattering of sunlight by atmospheric molecules and particles is detected with the Lidar as a sky background intensity, noted $\mathbf{I}_{sb}$ in equation (2). Geophysical factors contribute to $\mathbf{I}_{sb}$ such as the local meteorological conditions or the relative positioning between the Sun and the Earth. Sun sky scattering can drastically limit the range accessible to the perpendicular backscattering coefficient $\beta_{p,\perp}$ and induce photon noise.

**a. Polarization components of sunlight scattered by the atmosphere**

To minimize sky background contribution, a field stop and a daylight suppression band-pass filter are often inserted. In addition, we studied the polarization components of the sky background intensity vector $\mathbf{I}_{sb}$. These p and s-sunlight polarization components are defined with respect to the solar scattering plane, represented in figure 2a, together with the Lidar station (source and detector). The scattering angle is the solar zenith angle $\theta_s$, whose cosine is equal to $\cos(\theta_s) = \sin(\ell)\sin(\delta_s) + \cos(\ell)\cos(\delta_s)\cos(h)$, where $\ell$ is Lidar station latitude, $\delta_s$ is the solar declination angle and h is the local hour angle of the Sun. The p and s-polarization components of the sky intensity have been calculated by assuming a standard molecular atmosphere. In the presence of aerosols, the p-polarized component will increase (if these aerosols are spherical) or both polarized components will increase (if some aerosols are non-spherical). By assuming an unpolarized sunlight, the ratio between p and s-polarization components of $\mathbf{I}_{sb}$ can be expressed by using the molecular differential scattering cross-sections dependence on the scattering angle $\theta_s$ [28]:

$$\frac{I_{sb,p}}{I_{sb,s}} = \rho_0 + (1 - \rho_0)cos^2(\theta_s) \quad (10)$$

where $\rho_0$ is the depolarization factor of the standard molecular atmosphere [38]. Hence, the p-polarized component $I_{sb,p}$ is always below the s-component $I_{sb,s}$. We then projected these polarization components on the $\{//,\perp\}$-polarization Lidar axes by using figure 2a to obtain: $I_{sb,//} = \sin^2(h)I_{sb,p} + \cos^2(h)I_{sb,s}$ and $I_{sb,\perp} = \cos^2(h)I_{sb,p} + \sin^2(h)I_{sb,s}$. Hence, from sunrise to sunset, the two polarization sky background components cross twice during daytime.

**Please insert figure 2 here.**

**b. Experimental implications for measuring low depolarization ratios**

Figure 2b shows the measured sky background intensity $I_{sb}$ as detected on each {//, ⊥}−polarization channel at λ = 355 nm on July 3rd 2011. These observations agree with the above $I_{sb,//}$ and $I_{sb,\perp}$-expressions. In particular, when the sky background is at its maximum, $I_{sb,\perp}$ is below $I_{sb,//}$. Hence, to accurately measure depolarization ratios in the range of a few percents during daytime, it is interesting to match the perpendicular polarization sky background component with the perpendicular Lidar signal, which is approximately 100 times lower than the parallel Lidar signal. This polarization matching can be accomplished by rotating the laser linear polarization with a half-wave plate. In this situation, a new calibration is necessary (see section 3.4).

## 3. UV-VIS polarization Lidar experimental set-up

Lyon Lidar station (45.76 N, 4.83 E, France) is a home-built Lidar station, designed to remotely measure the polarization-resolved backscattering properties of tropospheric aerosols with a high spatial vertical resolution, a high sensitivity and a good accuracy. Hence, as shown in this section, our Lidar experimental set-up has been designed by carefully analyzing the role of each optical component on the spectral and polarization (λ, $\boldsymbol{\pi}$) optical properties of the photons backscattered from the troposphere. Polarization-resolved backscattering properties are studied at two wavelengths, in the UV and the VIS spectral range (usually referred to as 2β + 2δ-Lidar system in the Lidar community). As explained in the introduction, the choice for UV-light enables an increased sensitivity to fine particles [17]. In this section, we first describe our UV-VIS polarization-sensitive Lidar experimental set-up (emitter, receiver and detector). Then, by using section 2.2 theoretical considerations, we specify our detector by analyzing the specific role of each optical component through laboratory and field measurements.

### 3.1 The Lidar emitter and the receiver

The Lidar emitter and receiver are represented in figure 3 and the corresponding optical components are specified in Table 1. The laser head and the telescope are mounted on the same optical bench, kept free from vibration from the floor by buffers. The laser head is a doubled (VIS) and tripled (UV) Nd:YAG laser, emitting linearly polarized 10 ns duration laser pulses in the UV ($\lambda$ = 355 nm) and the VIS ($\lambda$ = 532 nm) spectral range with a 10 Hz repetition rate, for energies of approximately 10 mJ in the UV (20 mJ in the VIS spectral range). The laser head is fired for a sequence of 4000 laser shots by externally triggering the laser flash lamps. Then, each laser beam enter the emitter optics system, detailed in figure 3, composed of an emission polarizing beamsplitter cube (emission PBC), a half-wavelength plate ($\lambda/2$) and a ×2.5 beam expander (BE) to reduce the laser divergence down to 0.4 mrad while ensuring eye-safety. The emission PBC (Melles Griot, PBSO) improves the laser linear polarization rate to better than 10 000:1. The half-wavelength plate is used to adjust the laser linear polarization so that both wavelengths are emitted with the same linear vertical s-polarization. The 2$\lambda$-laser beams are then combined with a 2''-diameter dichroic mirror (DM, Melles Griot LD5644) which preserves the incident laser polarization ($T_s$(355 nm) > 99.5 %, $R_s$(532 nm) > 99.5 %) and then directed towards the atmosphere in the Eastward direction by an elliptical mirror ($M_E$) , also used for redirecting backscattered photons on a 200 mm diameter f/3-Newtonian telescope. During the alignment procedure, the telescope has been precisely positioned with respect to the laser beam axis, defined by two pinholes (see figure 3), by redirecting the 2$\lambda$-laser beams on the center of the telescope primary mirror by using two pentaprisms, as developed for precise alignment procedures (in the tens of micro-radian range) in quantum atom optics experiments [39]. We then identified the position of the telescope's focus as the intersection point of the 2$\lambda$-laser beams, originating from infinity to simulate backscattered photons from the atmosphere. The field of view (FOV) of the telescope − 2.5 mrad − is determined by a 3 mm-diameter pinhole inserted at the telescope's focus, and was chosen to minimize multiple scattering and solar sky background contributions to the Lidar signals. Moreover, the pinhole diameter was determined with the constraint to achieve lowest possible geometric compression, defined as the overlap function O(z) between the laser beam divergence and the receiver FOV. We numerically simulated the overlap function O(z) as a function of the laser initial diameter and beam divergence, the telescope's focal length, the primary and secondary mirror diameters and the pinhole diameter (there is no off-axis distance). With our 3 mm-diameter pinhole, the overlap function O(z) is equal to unity for z-altitudes above 150 meters above ground.

**Please insert figure 3 here.**

### 3.2 Lyon home-built UV-VIS Lidar polarization detector

The detector **D** is designed to efficiently separate backscattered photons with respect to their (λ, **π**)-spectral and polarization optical properties. The **D**-inside optical composition is represented in figure 4 through a top view and a 3D-exploded side view of each polarization channel. The specifications of the corresponding optical components are given in Table 2. **D** is housed in a small box, mounted on a rigid test bench located in the (x,y)-plane. Two 1 mm-diameter pinholes, located at the entrance and the detector exit, define the x-detector beam axis, which is materialized by a He-Ne laser, mounted on the detector bench. Use of a diffuser and observation of diffraction rings allowed defining the detector axis with a maximum deviation of 0.5 $mm.m^{-1}$, corresponding to 0.5 mrad. The He-Ne laser was used to position the detector at right angle with respect to the telescope axis by using a third pentaprism, hence merging the backscattered photons pathway with the detector beam axis (both axes being materialized by lasers) with better than 1 milliradian accuracy. Moreover, the He-Ne laser was used as an alignment laser, allowing **D** to be transportable, to allow mechanical alignment and precise optical specification of the inside detector at the laboratory, as detailed in section 3.3.

**Please insert figure 4 here.**

In between the two 1 mm-pinholes, backscattered photons are wavelength separated by using two dichroic beamsplitter ($DB_\lambda$), one for each λ-wavelength, which act as a low-pass filter selecting the desired UV,VIS wavelength. Each $DB_\lambda$ is positioned at 45° with respect to the detector x-axis and each λ-channel is polarization-resolved by using two successive polarizing beamsplitter cubes (PBC, see figure 4 exploded-view) which efficiently partitions the backscattered polarization $\boldsymbol{\pi}$. Sky background is reduced by a very selective band-pass interference filter ($IF_\lambda$) centered on the molecular Cabannes's line. The resulting molecular depolarization $\delta_m$ is hence slightly dependent on temperature variations: from 180 to 300 K, the error on $\delta_m$ is below 1 % [29]. Finally, (λ, $\boldsymbol{\pi}$)-backscattered photons are detected with a photomultiplier tube (PMT) having a 8 mm-diameter photocathode. The resulting four channel (λ, $\boldsymbol{\pi}$) photoelectrons are then sampled with two transient recorder (Licel, 12 bits, 20 and 40 MHz sample rate) and a two channels acquisition board (National Instruments MI, 12 bits, 50 MHz sample rate). A Labview program has been designed for externally triggering the laser head and recording the (λ,$\boldsymbol{\pi}$) range-resolved data acquisitions. Statistical error on the Lidar signals is reduced by operating acquisitions over 4000 laser shots, then performing high frequency filtering and range-averaging to lead to a final vertical resolution of 75 meters.

### 3.3 Specifying the Lidar performances for the polarization measurement

An ideal polarization Lidar has the ability to measure depolarization ratios with a high sensitivity (from a few to several tens of percents) and a high accuracy (by minimizing statistical and systematic errors). Statistical errors can be reduced by range and laser shots-averaging. In contrary, systematic errors lead to system bias that are crucial for depolarization ratios measurements in the range of a few percents, as described in section 2.2. In this paragraph, the performances of our home-built Lidar (emitter, receiver and detector) for sensitive and accurate dual-wavelength polarization measurements are specified through laboratory and field measurements. To fulfill the requirements derived from theoretical considerations, the (λ, **π**)-spectral and polarization properties of each optical component of the detector have been specified on a laboratory test bench. Manufacturer's specifications cannot be used, since these specifications are generally achieved for unpolarized continuous white light illumination, while the Lidar backscattered photons are polarized monochromatic pulses. This laboratory specification gives the evolution of a backscattered photon entering the detector **D** with (λ, **π**)-optical properties, throughout the detector box. A major achievement of this work is the derivation of the detector transfer matrix $[\mathbf{M_D(\lambda)}]$, which relates the intensity of backscattered photons between the detector entry and its exit, as a function of their (λ, **π**)-spectral and polarization optical properties. To our knowledge, such a specification is generally not reported in the literature. Here, our approach is different since we are interested in low atmospheric depolarization ratios, in the range of a few percents, which necessitates an accurate (λ, **π**)-detector specification.

**a. Specifying the Lidar emitter performances**

As a consequence of section 2.2.a, the polarization rate of the emitter unit must be carefully defined for accurate measurements of low depolarization ratios in the 1 %-range. We hence performed a laboratory experiment to control that the emitting optics (mirrors, dichroic beamsplitter) did not change the linear polarization state of the laser beam ($I_p/I_s$ > 10 000:1 after the emission PBC). To account for possible depolarization from the emission optics, which cannot be compensated by using a half-wave plate, we inserted a plane mirror above the mirror ($M_E$) to reflect the incident laser beam backwards down to the emission PBC. On the way back to the PBC, the ⊥-component of the laser beam, reflected by the emission PBC, can be analyzed with a supplementary PBC (not represented on figure 3). We checked that $R_p I_{//} >> R_s I_{\perp}$ (with $R_p = 1 - T_p$ and $R_s = 1 - T_s$ defined with respect to emission PBC). Hence, the cross-polarized component is negligible compared to the parallel polarized-component, ensuring negligible polarization dependent reflectivity of the emission mirrors.

**b. Specifying the Lidar detector performances**

**Detector mechanical specification**

In our home-built detector, each mechanical component has been fixed with respect to the others on a very rigid flat surface plate. During the detector building-up, to ensure precise mechanical alignment, the detector box was transported directed at the emission optics system exit: the detector axis was hence defined with respect to the laser beam axis, reported as a control with a thin quartz plate and a He-Ne laser (see figure 5). The verticality of the $DM_{\lambda}$, crucial for accurate polarization measurements (see section 2.2.d), has been adjusted with a one milliradian precision (by fixing each $DB_{\lambda}$ plate on a beam steering holder) and positioned at 45° with respect to the detector x-axis. Then, the PBC's were precisely positioned to ensure that both detected polarization components probe the same atmospheric volume. This crucial point has been checked through a laboratory experiment where both polarization channels counted the same photon flux for an incident 45° linear polarization. A similar experiment was conducted in the real atmosphere where the two polarization Lidar signals were equally populated when orienting the λ/2-emission plate at ± 45° with respect to the laser linear polarization. These mechanical specifications did not evolve for several months and the use of the 2λ-polarization Lidar did not reveal any mechanical drift or severe deterioration of the alignment.

**Specifying the $DB_\lambda$ dichroic beamsplitters**

Ideally, the dichroic beamsplitters must reflect photons at the λ-wavelength while preserving their polarization. In the literature, negligible attenuation of the optics before the PBC is often assumed [16]. We have measured the $R_p$, $R_s$-reflectivity coefficients of each $DB_\lambda$ plate (defined with respect to the $DB_\lambda$ incidence plane) on a dedicated laboratory test. As shown in figure 5 (top), our test bench is composed of a two lenses optical system, having an f/3 numerical aperture to simulate collection of backscattered light from the atmosphere by the f/3 Lidar's telescope. Table 2 presents the measured (λ, $\boldsymbol{\pi}$) reflectivity for each $DM_\lambda$-plate. Since $R_p$ differs from $R_s$, the $DM_\lambda$ vertical positioning achieved during the mechanical alignment procedure is effectively crucial, as a consequence of section 2.2.d theoretical development.

**Please insert figure 5 here.**

**Specifying the polarization PBC, polarization cross-talks**

As developed by J. Alvarez et al. [36], the insertion of a second polarizing beamsplitter cube (PBC) ensures the polarization purity of the perpendicular polarization channel. In the UV (resp. VIS) spectral range, manufacturers specifications (CVI Melles Griot) indicate $T_p > 90.0$ % (resp. > 95.0 %) and $R_s > 99.0$ % (resp. 99.9 %) for a 2° field of view ensured by our mechanical alignment. We have measured the $T_p$, $R_s$-coefficients of each PBC on the dedicated laboratory test bench presented in figure 5 (bottom), where the polarization state of the incident light is controlled by a λ/2 plate. The VIS-PBC's exhibit $T_p$ and $R_s$-values very close to 1 so that polarization cross-talk is fully negligible at λ = 532 nm. Table 2 presents the measured $T_p$, $R_s$-coefficients for each UV-PBC. $PBC_1$, which exhibited the highest $R_s$-value, has been used as shared PBC, to improve the detected perpendicular intensity, while $PBC_2$ (resp. $PBC_3$) was inserted on the parallel (resp. perpendicular) channel. Hence, our measured polarization cross-talks coefficients are $CT_{//} = R_{p,1}T_{s,3} = 4 \times 10^{-6}$ and $CT_{\perp} = T_{s,1}T_{s,2} = 0$.

**Optimizing the detection of the backscattered photons flux**

The spectral separation of our UV-VIS detector is not perfect, due to the limited spectral rejection of the dichroic beamsplitters, which are built for efficient operation in either the UV or the VIS spectral range. However, in our detector, this contamination, measured with the laboratory test bench, is completely negligible thanks to the use of very selective band-pass interference filters (OD 5 at $\lambda$ = 532 nm for the $IF_{UV}$ and at $\lambda$ = 355 nm for the $IF_{VIS}$). Then, using our f/3 optical test bench, we visualized the light pathway throughout the detector to prevent any eventual light blocking within the telescope's FOV. Moreover, the parallel (perpendicular) polarized intensity $I_{//}$ ($I_{\perp}$) has been detected in (at right angle to) the plane of the detector, while in the literature [16], the //-polarized signal is often detected on the s-branch of the PBC since $R_s > T_p$. Since the difference between $R_s$ and $T_p$ is very small, as explained in section 2.3, we matched the $\perp$-polarized Lidar signal, which is low and hence difficult to accurately measure, with the lowest polarization component of the sky background intensity (i.e. the $\perp$-polarized sky background intensity around midday). Finally, to improve the Lidar signals quality, the position of the 8 mm-photocathode PMT has been optimized with respect to lens ($L_2$). No photon counting was hence necessary, despite the very low Lidar $I_{\perp}$-intensity for low depolarization ratios measurements.

**Detector transfer matrix**

The performances of our spectral and polarization-resolved detector can be summarized by writing the detector transfer matrix $[\boldsymbol{\eta}(\boldsymbol{\lambda})]$ corresponding to the $\lambda$-detected wavelength, which relates the backscattered photons intensity vector $\mathbf{I} = [I_p, I_s]^T$ to the detected backscattered photons intensity vector $\mathbf{I}^* = [I^*_{//}, I^*_{\perp}]^T$ (see equation (3)). In this formalism, the role of the dichroic beamsplitter $DB_\lambda$ is summarized by a matrix $\mathbf{M_{DB}}$, detailed in the appendix, which is diagonal since the off-set angle $\theta_0$ is null. Likewise, the $\boldsymbol{\pi}$-separation achieved by the PBC's is taken into account with a diagonal matrix (with better than $5 \times 10^{-6}$ accuracy in the UV-spectral range). By noting that the p-component of the $DB_\lambda$ corresponds to the s-polarized component of the PBC, we get:

$$[\boldsymbol{\eta}_{(UV)}] = \begin{bmatrix} g_{//,UV} \\ g_{\perp,UV} \end{bmatrix} \begin{bmatrix} 0.99 & 0 \\ 4 \times 10^{-8} & 0.72 \end{bmatrix} \text{ and } [\boldsymbol{\eta}_{(VIS)}] = \begin{bmatrix} g_{//,VIS} \\ g_{\perp,VIS} \end{bmatrix} \begin{bmatrix} 0.87 & 0 \\ 0 & 1 \end{bmatrix} \quad (11)$$

where $g_{//,\lambda}$ and $g_{\perp,\lambda}$ are the gains resulting from the applied PMT voltages on the λ-polarization channel. Then, for our dual-wavelength polarization detector, a detector transfer matrix $\mathbf{M_D}$ can be written to relate the backscattered photons 2λ-intensity vector $\mathbf{I_{2\lambda}} = [I_{UV,p}, I_{UV,s}, I_{VIS,p}, I_{VIS,s}]^T$ to the detected backscattered photons 2λ-intensity vector $\mathbf{I^*_{2\lambda}} = [I^*_{UV,//}, I^*_{UV,\perp}, I^*_{VIS,//}, I^*_{VIS,\perp}]^T$:

$$\mathbf{I^*_{2\lambda}} = [\boldsymbol{M_D}]\mathbf{I_{2\lambda}} \text{ with } [\boldsymbol{M_D}] = \begin{bmatrix} [\boldsymbol{\eta}_{(UV)}] & \begin{matrix} 0 & 0 \\ 0 & 0 \end{matrix} \\ \begin{matrix} 0 & 0 \\ 0 & 0 \end{matrix} & [\boldsymbol{\eta}_{(VIS)}] \end{bmatrix} \quad (12)$$

to underline the spectral selectivity of our detector (no contamination between UV and VIS channels). As a conclusion, our (λ, $\boldsymbol{\pi}$)-optical detector is characterized by a bloc diagonal $\mathbf{M_D}$-matrix (with $4 \times 10^{-8}$ accuracy) underlying that our detector efficiently partitions backscattered photons as a function of their polarization $\boldsymbol{\pi}$ (as a consequence of the use of two PBC) and their wavelength λ (as a consequence of OD 5-interference filters which ensure efficient wavelength separation without contamination).

## Multiple scattering effects

Multiple scattering processes may induce some depolarization, even in the presence of spherical scatterers. Hence, to relate the atmosphere depolarization ratio to particles non-sphericity, we accounted for possible multiple scattering effects. The contribution of multiple scattering to the measured depolarization ratio $\delta$ has been extensively studied in the literature [40,41]. In the case of optically dense objects in the PBL, the depolarization ratio varies as an exponential law of the telescope's FOV [42]. We have measured the atmosphere depolarization ratio $\delta$ as a function of the receiver telescope's FOV. As shown by Tatarov et al. [42], for FOV below 8 mrad, the depolarization ratio is almost independent on the receiving FOV, and hence exclusively due to particles non-sphericity. Hence, our current 2.5 mrad receiving FOV, obtained by inserting a 3 mm-diameter pinhole at the telescope's focus, is too small to provide depolarization via multiple scattering processes.

### 3.4 Calibration procedure for the polarization measurement

Since the detector transfer matrix is diagonal, depolarization ratios are known with a multiplicative constant that must be accurately determined for quantitative depolarization ratios measurements. This multiplicative factor depends on the reflectivity and the transmission of the dichroic beamsplitter and the PBC's, the $IF_\lambda$-transmission and the gain of the PMT at the applied voltages. Since the relation between the PMT-gain and the applied voltage is not precisely known, a calibration procedure is necessary to determine the calibration constant corresponding to the $\lambda$-channel, hereafter referred to as the electro-optics calibration constant $G_\lambda$. As shown by J. Alvarez [36], in the absence of misalignment between the laser linear polarization and the parallel axis of the detector PBC, the measured depolarization ratio $\delta^*$ is related to the atmosphere depolarization $\delta$ by the simple relation $\delta^* = G_\lambda\delta$, as detailed in equation (8). This section presents our experimental determination of $G_\lambda$ for the two 355 and 532 nm polarization channels by using Alvarez's method [36]. To retrieve $G_\lambda$, we introduced a controlled offset angle $\varphi$ between the laser linear polarization and the parallel axis of the detector's PBC [36]. In this case, the behavior of $\delta$ as a function of $\delta_0$ depends on the $\varphi$-angle with a law that is given in equation (8). These controlled amounts of polarization cross-talks can be used to retrieve $G_\lambda$ by adjusting $\delta$ as a function of the $\varphi$ calibration angle under stable atmospheric conditions. Other methods [16,29] are applicable but molecular calibration is of limited accuracy since a particle-free atmosphere does not rigorously exist. Setting $\varphi$ to ±45° leads to the same backscattered intensity on each polarization channel, so that $G_\lambda$ is then simply equal to $\delta^*/\delta$ [16]. The accuracy of this ± 45° calibration method is adequate but limited by the possibility to have an exact 90° rotation control and by possible PMT-saturation (for $\delta$ = 1 %, $I_\perp$ is multiply by 50). Figure 6 presents the results of our calibration procedure, obtained by applying Alvarez's method. Two calibration curves are provided for backscattering at 355 nm and at 532 nm. The $G_\lambda$-value is retrieved by adjusting the measured points with their error bar by using equation (8). We hence obtained $G_{UV} = 29.16 \pm 0.22$ and $G_{VIS} = 16.69 \pm 0.23$, the accuracy originating from the use of several $\varphi$-calibration angles. With this method, the maximum $\varphi$-angle value used is 23°, which limits possible PMT-saturation effects (this time, for $\delta$ = 1%, $I_\perp$ is multiplied by 17 which respects the PMT linearity). Hence, $G_\lambda$ is known with better than 2 %-accuracy.

**Please insert figure 6 here.**

## 4. Application to free troposphere and urban aerosols remote sensing

In this section, our dual wavelength polarization detector is used to remotely measure the polarization backscattering properties of tropospheric aerosols at Lyon. Hence, the sensitivity and the accuracy of our detector for ($\beta_p$, $\delta_p$)-measurements are evaluated under atmospheric conditions, as a consequence of the theoretical considerations derived in section 2 and the specification of our detector achieved in section 3.

High particles depolarization ratios have been measured in the Lyon troposphere during volcanic and dust episodes that occurred at Lyon during the last two years, as a consequence of the highly-irregularly shape of volcanic ash and Saharan dust particles. These events have been an opportunity to test the ability of our detector to measure high UV-particle depolarization ratios, as high as (40.5 ± 8.0) % for volcanic ash particles at 4 kilometers altitude [22], or $\delta_p$ = (19.5 ± 3.5) % for Saharan dust particles [24]. The achieved UV-sensitivity, specified in section 3, allowed distinguishing different volcanic layers, having different depolarization ratios, with 75 meters-vertical resolution [21]. However, no measurements were made in the VIS spectral range.

As explained in the introduction, we here focus on the measurement of low depolarization ratios, in the range of few percents, which is the most frequently observed situation in the Lyon troposphere. In figure 7, we present a time-altitude map, showing the spatial and temporal evolution of the optical properties of Lyon tropospheric particles, on 18th October 2011 between 13h30 and 18h00 up to 4 kilometers altitude. The plotted optical properties are the parallel and perpendicular particle backscattering coefficient $\beta_{p,//}$ and $\beta_{p,\perp}$ and the corresponding particle depolarization ratio $\delta_p$, in both the UV and the VIS spectral range. To facilitate the interpretation of these data, the color scales have been adjusted in each color plot, to put light on the achieved sensitivity. To our knowledge, the calibrated UV-particle depolarization $\delta_p$-map, achieved for tropospheric urban particles, is a new achievement. These plots show that the observed tropospheric particles are mainly distributed into two major distinct atmospheric layers: the PBL (for altitudes below 1 km), outlined by a strong temperature inversion observed at 1 km altitude, and the free troposphere (for altitudes above 1 km). In the PBL, the layer is relatively homogeneous, while in the free troposphere, two secondary layers (around 2 and 3 km altitude), with temporal non homogeneity, can be distinguished.

In the PBL, the $\beta_{p,//}$-values are high in the UV, in the range of $5 \times 10^{-6}$ $m^{-1}.sr^{-1}$, a value usually observed during a smog episode [25]. Within our error bars (to be seen in figure 8), the $\beta_{p,//}$-values are significantly lower than in the VIS. This behavior can be explained by considering that backscattering in the UV more efficiently addresses ultra-fine and fine particles than backscattering in the VIS spectral range [17]. Therefore, in the PBL, particles are mostly distributed in the fine particle mode of the particles size distribution, with radii typically around 300 nm [43]. In the free troposphere, the VIS $\beta_{p,//}$-values are often comparable to the UV $\beta_{p,//}$-values, so that both fine and coarse particles modes are addressed.

While the $\beta_{p,//}$-Lidar channel is sensitive to both spherical and non-spherical particles, the $\beta_{p,\perp}$-Lidar channel is non-spherical particles specific [24]. Hence, the $\beta_{p,\perp}$-map provides the spatial and temporal distribution of tropospheric non-spherical particles. The sensitivity achieved in section 3 enables to measure very small $\beta_{p,\perp}$-values, as low as $3 \times 10^{-8}$ $m^{-1}.sr^{-1}$. Therefore, in the PBL, for altitudes below 500 meters, even non-spherical coarse particles can be detected in the VIS. Due to their inertia, these particles seem unable to reach higher altitudes by convection. In the meanwhile, in the UV, the $\beta_{p,\perp}$-map exhibits non-spherical fine particles (more efficiently detected on the UV-polarization channels), following the convective atmospheric movements up to the PBL, after a wind speed change occurred at 15 h. Hence, despite the very low measured depolarization ratios, the dynamics of non-spherical particles is retraced. Only a few percent particle depolarization ratio are measured in the UV, since $\beta_{p,//}$ is very high due to the smog episode. In the VIS, $\delta_p$ is higher as a consequence of low concentration coarse mode particles. In the free troposphere, in the VIS, non-spherical particles lay into two relatively homogeneous major layers having different thicknesses. The different distribution observed in the UV indicates that the non-spherical particles size distribution is non homogeneous in the free troposphere. The corresponding $\delta_p$-measurements exhibit particle depolarization ratios between 6 and 10 %. The meteorological analysis enables to think that the corresponding air masses originate from the Atlantic region and contain non-spherical particles, such as sea-salt particles, as confirmed by laboratory measurements which reveal an intrinsic depolarization ratio in the 10 %-range [44].

**Please insert figure 7 here.**

Vertical profiles of $\beta_{p,//}$, $\beta_{p,\perp}$ and $\delta_p$ are proposed in figure 8 in the UV and the VIS at 14h45 to provide the error bar on each measured coefficient, hence addressing our achieved sensitivity and accuracy. As observed during tropospheric volcanic ash events [21] or in the stratosphere [30], scattering does not necessarily correlate with depolarization, since $\delta$ does not follow the superposition principle as described in section 2.1. Vertical profiles are limited up to 4 km to preserve a high signal-to-noise range ratio. The proposed error bars result from precise $R_{//}$-evaluation and accurate $\delta$-measurement. Within our error bars, at 14h45, particles in the PBL are more efficiently detected in the UV-channel, indicating that these particles are preferably fine low depolarizing particles. In the free troposphere, the depolarization behavior results from a complex mixing of fine and coarse particles. Despite strong UV-molecular scattering, our error bars are very low as a consequence of our very precise calibration procedure and our laboratory detector building up, optimization and specification. The relative error on the $\beta_p$-coefficient does not exceed 10 % while the maximum error on the particle depolarization ratio $\delta_p$ is 23 %, calculated by using equation (5). To our knowledge, such values have never been reported in the literature, especially in the UV spectral range. Moreover, very low depolarization ratios, as low as only a few percents, are measured with accuracy. In the UV (resp. VIS), at z = 800 m, we measured $\delta_p$ (UV) = (4.2 $\pm$ 0.3) % (resp. $\delta_p$ (VIS) = (3.4 $\pm$ 0.3) % at 14h45. Hence, our detection limit is 2 × 0.3 % = 0.6 %, a value comparable to the molecular depolarization. As a conclusion, the particle depolarization in the PBL should be considered as different from zero.

**Please insert figure 8 here.**

## 5. Conclusions

In this paper, a dual-wavelength polarization Lidar detector has been built, optimized, specified and operated, by starting from the very beginning. Our new detector provides remote measurements of the polarization-resolved backscattering properties of tropospheric aerosols, in the UV (at 355 nm) and the VIS (at 532 nm) spectral range, with a high spatial vertical resolution, a high sensitivity and a reliable accuracy. To our knowledge for the first time, a calibrated particle depolarization $\delta_p$-map has been achieved in the UV spectral range for tropospheric particles, despite strong molecular scattering.

We first identified the relevant parameters for measuring low particle depolarization ratios, in the range of a few percents, from a theoretical point of view (see section 2). To trustworthy evaluate such depolarization ratios, it is necessary to evaluate the different system biases altering the backscattered polarization. In particular, the sensitivity of each bias for low depolarization measurements has been quantified in section 2. Then, in section 3, the spectral and polarization properties of our dua-wavelength polarization detector have been specified on a laboratory dedicated test bench, to satisfy the section 2 identified requirements. The role of the dichroic beamsplitter used for dual-wavelengths measurements, has been precisely addressed. Moreover, the backscattered photons flux has been optimized and the detector specifications have been reported in a synthetic blog-diagonal detector transfer matrix, underlying the partitioning efficiency of backscattered photons as a function of their polarization $\boldsymbol{\pi}$ (as a consequence of the use of two PBC) and their wavelength $\lambda$ (as a consequence of very selective IF ensuring efficient wavelength separation without contamination). After accurate polarization calibration procedure, we tested the sensitivity and accuracy of our ($\lambda$, $\boldsymbol{\pi}$)-Lidar detector under real atmospheric conditions by measuring particles backscattering coefficient ($\beta_p$) and depolarization ratio ($\delta_p$) for tropospheric aerosols. The $\beta_p$- coefficient, derived from the Klett's algorithm, has been calculated from an extinction-to-backscatter ratio $S_p$, numerically evaluated as a function of the PBL-thermodynamics by using a three-mode aerosols size distribution detailed in [25]. Accurate Raman Lidar measurements or HSRL can be an alternative methodology to derive $\beta_p$ [45]. The polarization detector measures UV-particle depolarization ratios over almost two orders of magnitude, from 0.6 % (detection limit very close to the molecular depolarization), up to 40 %, as observed during volcanic ash episodes. Such depolarization ratios are remotely measured over 4 kilometers, with a vertical range resolution of only 75 meters. The achieved sensitivity and accuracy enable to precisely retrace the polarization and backscattering properties of tropospheric aerosols, even in the presence of low depolarizing particles. Hence, conclusions on atmospheric physics can be drawn. The observed $\beta_p$ and $\delta_p$- time-altitude maps exhibit a different behavior in the UV and the VIS spectral range, as a consequence of the higher scattering efficiency of fine particles in the UV [17]. Hence, fine and coarse particles are addressed in the PBL (where a smog episode is observed) and in the free troposphere (where sea-salt particles are to be seen) with our dual-wavelength polarization Lidar. Spectroscopy of nano-sized atmospheric particles can then be remotely achieved.

As a conclusion, achieving sensitive and accurate low depolarization ratios measurement is difficult, especially in the UV spectral range where molecular scattering is strong. This difficulty obliged us to precisely analyze the relevant parameters for trustworthy measure particle depolarization ratios. Consequently, a major achievement of this work is the observation of non-spherical tropospheric particles in the PBL, in the UV and the VIS spectral range. This dual-wavelength particle depolarization ratio measurement may open new insights for further use in retrieval schemes aimed at deriving the particles microphysics. Knowledge on the solid-state content of the atmosphere may enable to explore new pathways in atmospheric photo-chemistry, especially for photo-catalytically heterogeneous reactions occurring at the PM surface [46]. In this context, knowledge of the particle linear depolarization ratio $\delta_p$ at two wavelengths, namely the UV and the VIS, is essential, as detailed in several theoretical publications [47,48], provided that sensitive and accurate Lidar polarization measurements are achieved.

## 6. Acknowledgments

The authors thank Marc Néri for his help in fine mechanics and Région Rhône-Alpes for research grant.

## 7. Appendix

In this appendix, we investigate the effect of a misalignment of the dichroic beamsplitter on the measured depolarization ratio $\delta^*$. To parameterize the magnitude and the direction of this misalignment, we introduce an offset angle $\theta_0$ defined in figure 1.d as the angle between the parallel laser linear polarization and the p-axis of the dichroic beamsplitter (defined with respect to the dichroic beamsplitter plane of incidence). The aim of this appendix is to derive the relationship between the measured depolarization $\delta^*$ and the atmosphere depolarization $\delta$ as a function of the $\theta_0$ offset angle and the $R_p$, $R_s$-reflectivity coefficients of the dichroic beamsplitter, hence justifying equation (9).

The incident electric field $\mathbf{E_i}$ on the dichroic beamsplitter can be written in the two involved mathematical bases, namely the (//, ⊥)-Lidar polarization basis and the (p,s)-dichroic beamsplitter basis. As shown by figure 1.d, a $\theta_0$-rotation angle enables to change from one basis to the other. We projected the incident electric field vector $\mathbf{E_i}$ of backscattered photons on the (p, s)-polarization basis to express the electric field vector $\mathbf{E_r}$ of the reflected wave:

$$\begin{bmatrix} E_{r,//} \\ E_{r,\perp} \end{bmatrix} = \begin{bmatrix} r_p \cos(\theta_0) & -r_s \sin(\theta_0) \\ r_p \sin(\theta_0) & r_s \cos(\theta_0) \end{bmatrix} \begin{bmatrix} E_{i,p} \\ E_{i,s} \end{bmatrix} \quad \text{(A-1)}$$

In this expression, we have introduced amplitude field reflectivity coefficients $r_p$ and $r_s$ defined as $r_p = E_{r,p}/E_{i,p}$ and $r_s = E_{r,s} / E_{i,s}$ where $E_{i,p}$ and $E_{i,s}$ are the components of $\mathbf{E_i}$ in the (p,s)-dichroic beamsplitter basis (the same notations are used for the reflected field $\mathbf{E_r}$). Then, by projecting the incident electric field in the (//,⊥)-polarization basis, equation (A.1) becomes:

$$\mathbf{E_r} = [\mathbf{m_{DB}}]\ \mathbf{E_i} \quad \text{with} \quad \mathbf{m_{DB}} = \begin{bmatrix} b - a sin^2\theta_0 & a cos\theta_0 sin\theta_0 \\ a cos\theta_0 sin\theta_0 & b - a cos^2\theta_0 \end{bmatrix} \qquad \text{(A-2)}$$

where the $\mathbf{m_{DB}}$-matrix relates the incident and reflected electric fields in the (//,⊥)-polarization basis and the two coefficients $a = r_p - r_s = \sqrt{R_p} - \sqrt{R_s}$ and $b = r_p = \sqrt{R_p}$ are determined by the dichroic beamsplitter $R_p$, $R_s$-reflectivity coefficients. Hence, reflection (or symmetrically transmission) on the dichroic beamsplitter induces a rotation of the linear polarization state of the light. In the ideal case, the dichroic beamsplitter is vertical, so that the p-axis is horizontal and $\theta_0$ is zero. If we exchange the // and ⊥-polarization channels, $\theta_0$ is then $\pi/2$. In both cases ($\theta_0 = 0$ or $\pi/2$), the $\mathbf{m_{DB}}$-matrix is diagonal so that no cross-talk is induced. To derive the measured depolarization ratio $\delta^*$ as a function of $\delta$, we now introduce intensities, proportional to the square of the electric field. Hence, equation (A.2) can be written for laser intensities vectors $\mathbf{I_r}$ and $\mathbf{I_i}$. By removing proportionality constants (which disappear in the $\delta^*$-calculation), we get:

$$\mathbf{I_r} = [\mathbf{m_{DB}}]\ \mathbf{I_i} \quad \text{with} \quad \mathbf{M_{DB}} = \begin{bmatrix} (b - a sin^2\theta_0)^2 & a^2 cos^2\theta_0 sin^2\theta_0 \\ a^2 cos^2\theta_0 sin^2\theta_0 & (b - a cos^2\theta_0)^2 \end{bmatrix} \qquad \text{(A-3)}$$

by noting that the (//,⊥)-polarization basis is orthogonal. As expected, the $\mathbf{M_{DB}}$-matrix is diagonal in the absence of offset angle $\theta_0$ (i.e. if $\theta_0 = 0$ or $\pi/2$). By noting that $\delta^* = I_{r,\perp}/I_{r,//}$ while $\delta = I_{i,\perp}/I_{i,//}$, we get the following relationship between $\delta$, $\delta_0$ and $\theta_0$, which is identical to equation (9):

$$\delta^* = \frac{a^2cos^2\theta_0 sin^2\theta_0 + \delta_0(b - acos^2\theta_0)^2}{(b - asin^2\theta_0)^2 + \delta_0\, a^2cos^2\theta_0 sin^2\theta_0} \qquad \text{(A.4)}$$

where the two coefficients a = $r_p - r_s = \sqrt{R_p} - \sqrt{R_s}$ and b = $r_p = \sqrt{R_p}$ are determined by the dichroic beamsplitter $R_p$, $R_s$-reflectivity coefficients.

**List of figures and table captions**

**Fig.1** System bias affecting the dual-wavelength polarization Lidar measurement: (a) presence of a small unpolarized component in the emitted laser beam, (b): imperfect separation of polarization components, (c): misalignment between the transmitter and receiver polarization axes, (d): role of the dichroic beamsplitter introduced for dual-wavelength detection. Top schemes represent the studied system bias while bottom graphs present the relative error on $\delta$ for different values of the bias parameter ($\varepsilon$, $CT_{//}$ and $CT_{\perp}$, $\varphi$ and $\theta_0$).

**Fig.2** Sky background contribution to the Lidar intensity. (a) Sun scattering plane geometry and orientation with respect to the Lidar laser source and the detector polarization $\{//,\perp\}$-axes. The emission laser is oriented to the East, and the angle between the solar scattering plane and the East is $\pi/2 - h$. (b) Measured sky background intensity $I_{sb}$ on each polarization $\{//,\perp\}$-axis as a function of the solar local angle on July 3rd 2011 at Lyon.

**Fig.3** Top view of the Lidar station with the laser head, the emitting optics (detailed in the dashed below box), the receiving optics (elliptical mirror ($M_E$), telescope) and the Lidar detector D. The laser beam is emitted vertically, along the z-altitude axis.

**Fig.4** (a) Top view of our home-built UV-VIS polarization Lidar detector D. (b) 3D-exploded side view of each polarization channel (UV, VIS) composed of two PBC's , one $IF_\lambda$ and a PMT.

**Fig.5** Laboratory test-bench with numerical aperture f/3 to simulate backscattered photons from the atmosphere. The top scheme is used for measuring the $DB_\lambda$ reflectivity while the bottom scheme enables the $T_p$, $R_s$-measurements of the PBC's polarization properties, using the $\lambda/2$ plate to control the incident polarization.

**Fig.6** Calibration curves obtained in the UV and the VIS with corresponding residue plot.

**Fig.7** Time-altitude maps of the parallel and perpendicular particle backscattering coefficient $\beta_{p,//}$ and $\beta_{p,\perp}$ and the corresponding particle depolarization ratio $\delta_p$, in the UV and the visible spectral range on October 18th 2011 at Lyon between 13h30 and 18h. The color scales have been adjusted on each map to enhance the achieved sensitivity in the UV and in the VIS. In between each 4000 laser shots-vertical profile, the laser has been shut down during 4 minutes.

**Fig.8** Vertical profiles of $\beta_{p,//}$, $\beta_{p,\perp}$ and $\delta_p$ on October 18th 2011 at 14h45 at Lyon in the UV (blue) and the VIS (green). Error bars on $\beta_{p,//}$ are calculated by using the maximum and minimum values of $S_p$ in the Klett's algorithm. Error bars on $\beta_{p,\perp}$ are derived from the section 2-derived relation : $\beta_{p,\perp} = (R_{//}\delta - \delta_m) \times \beta_{m,//}$ while error bars on $\delta_p$ are calculated by applying equation (5).

**Tab. 1** Optical specifications of the emission laser, the emitter optics and the Lidar receiver.

**Tab. 2** Optical specifications of the UV-VIS polarization detector D.

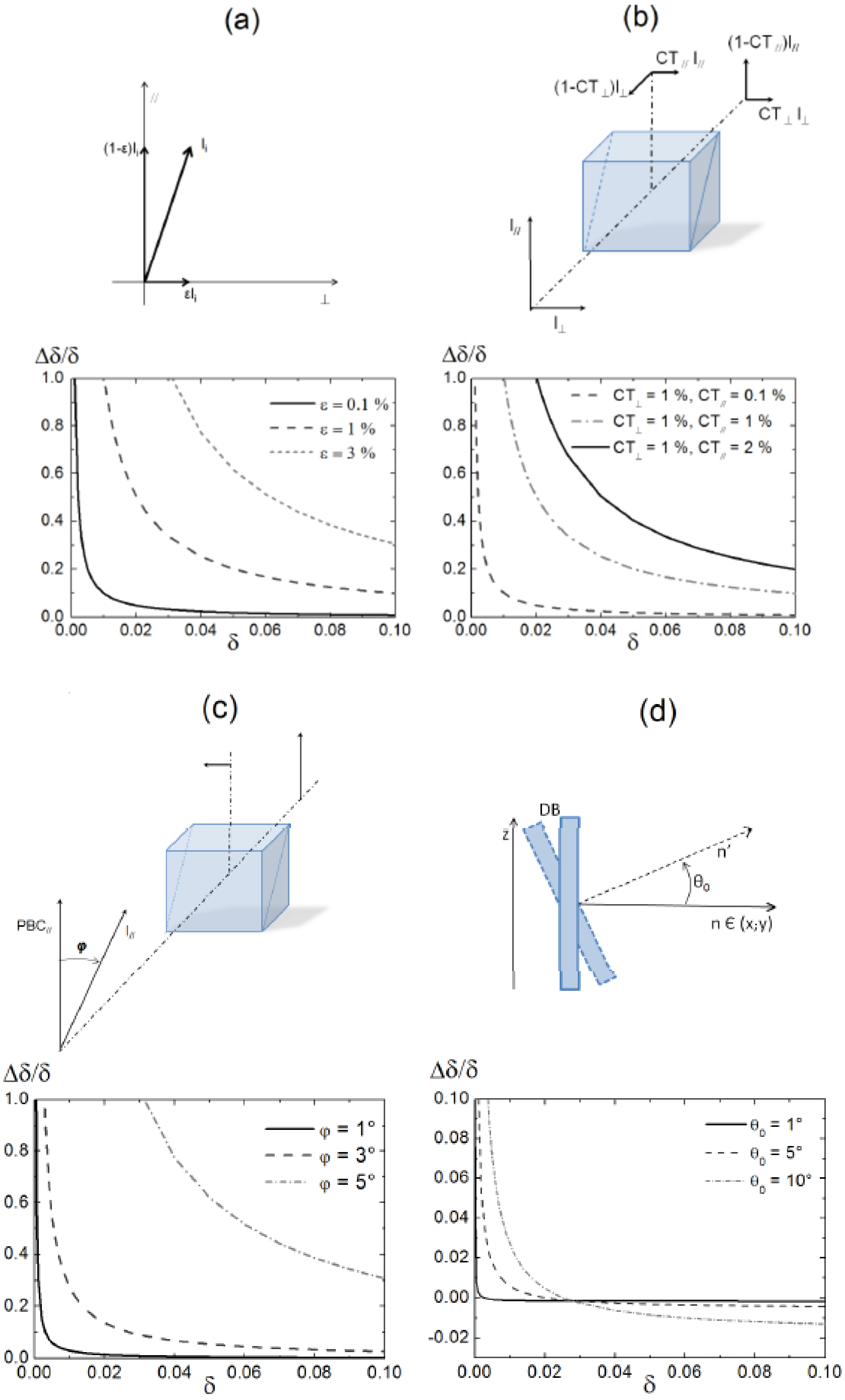

**Fig.1** System bias affecting the dual-wavelength polarization Lidar measurement: (a) presence of a small unpolarized component in the emitted laser beam, (b): imperfect separation of polarization components, (c): misalignment between the transmitter and receiver polarization axes, (d): role of the dichroic beamsplitter introduced for dual-wavelength detection. Top schemes represent the studied system bias while bottom graphs present the relative error on δ for different values of the bias parameter (ε, $CT_{//}$ and $CT_{\perp}$, φ and $\theta_0$).

**(a)**

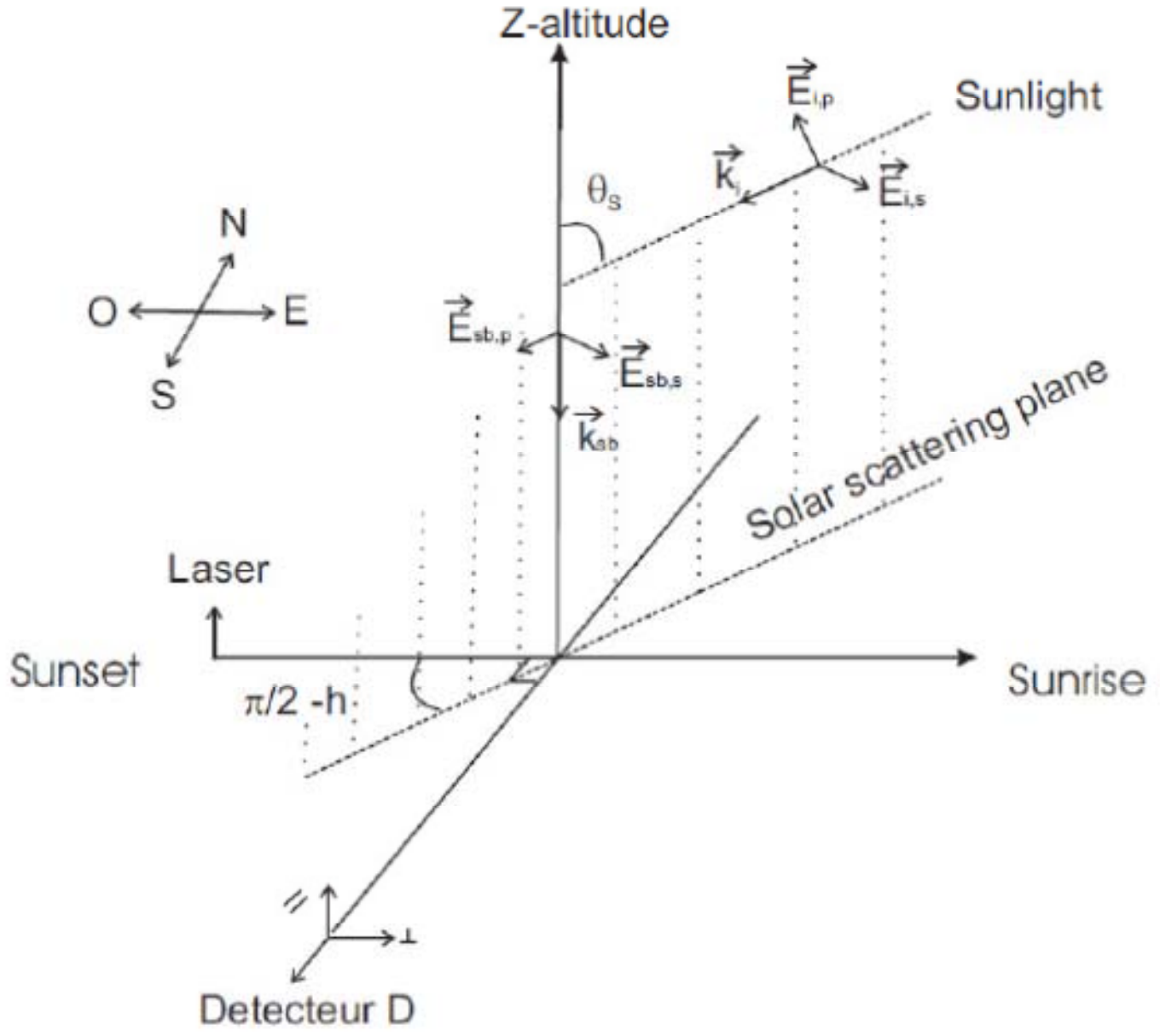

**(b)**

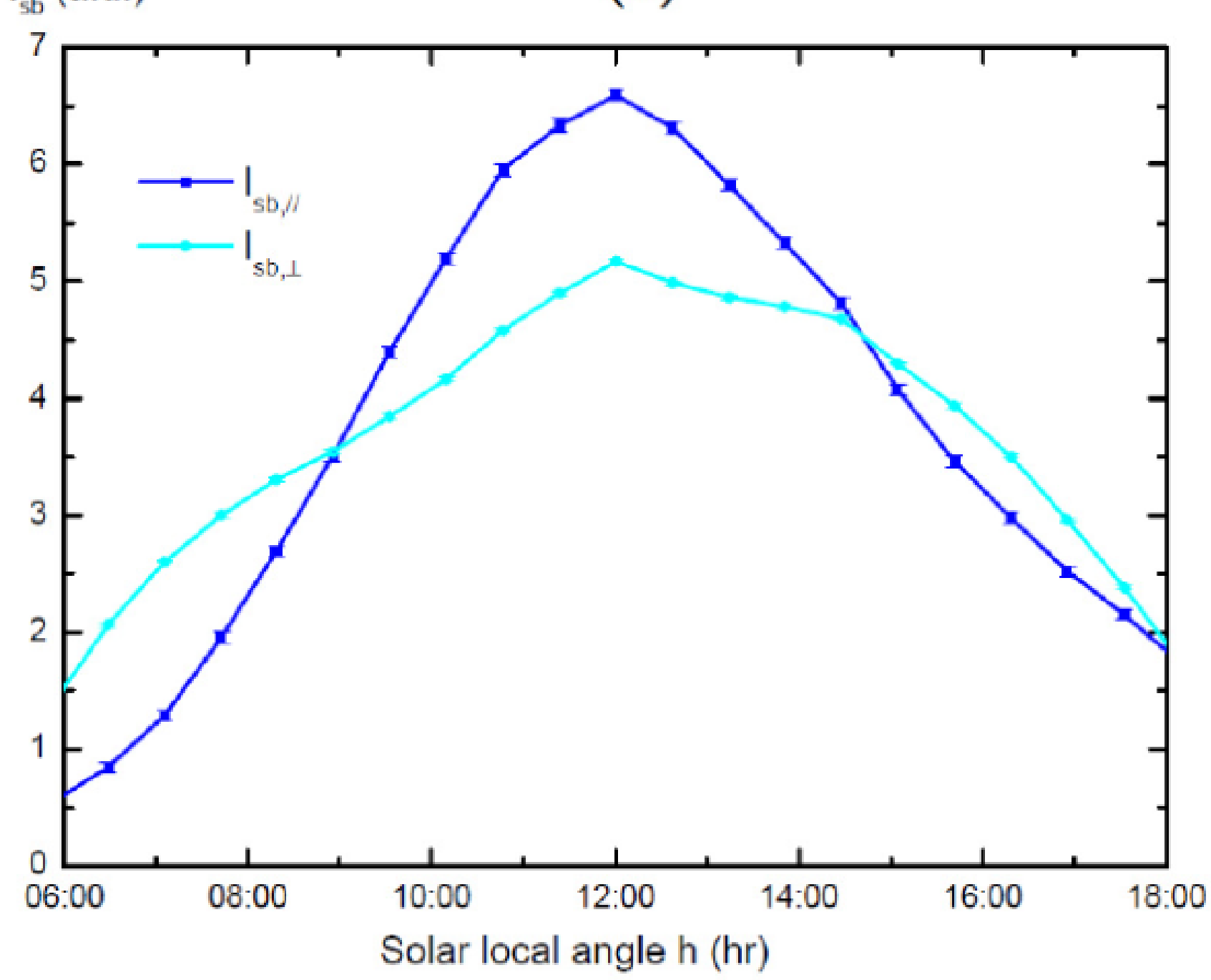

**Fig.2** Sky background contribution to the Lidar intensity. (a) Sun scattering plane geometry and orientation with respect to the Lidar laser source and the detector polarization {//,⊥}-axes. The emission laser is oriented to the East, and the angle between the solar scattering plane and the East is $\pi/2 - h$. (b) Measured sky background intensity $I_{sb}$ on each polarization {//,⊥}-axis as a function of the solar local angle on July 3rd 2011 at Lyon.

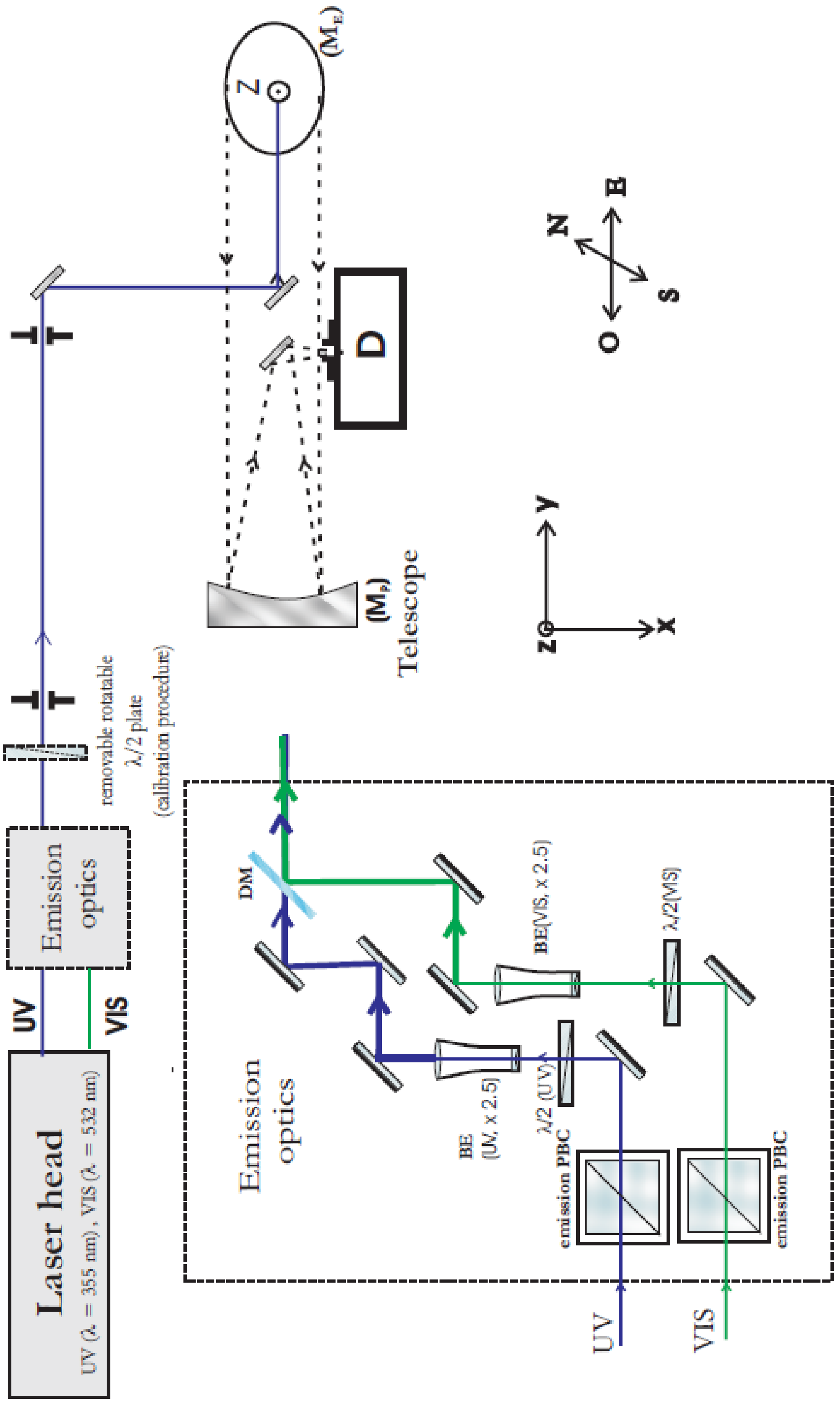

**Fig.3** Top view of the Lidar station with the laser head, the emitting optics (detailed in the dashed below box), the receiving optics (elliptical mirror ($M_E$), telescope) and the Lidar detector D. The laser beam is emitted vertically, along the z-altitude axis.

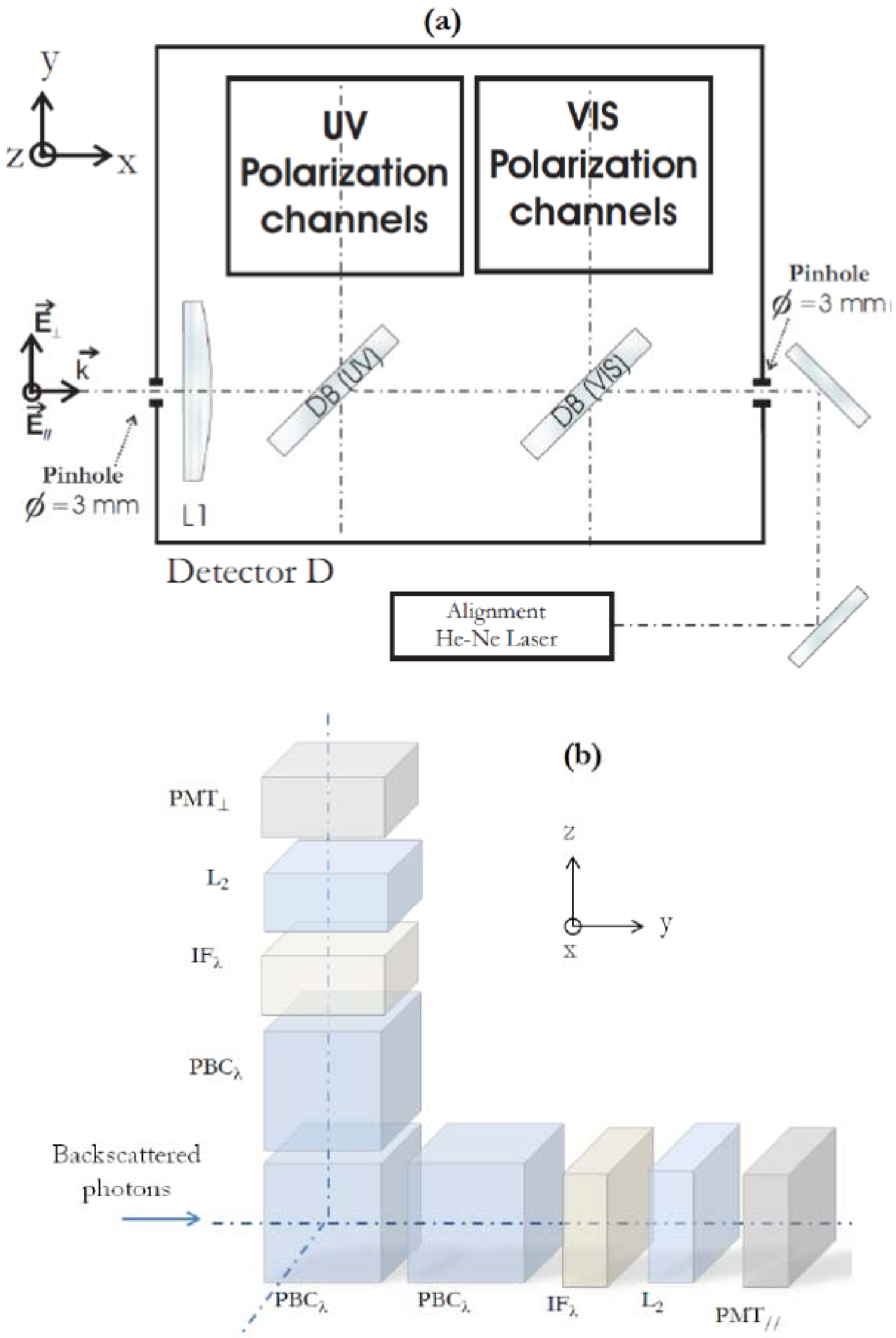

**Fig.4** (a) Top view of our home-built UV-VIS polarization Lidar detector D. (b) 3D-exploded side view of each polarization channel (UV, VIS) composed of two PBC's , one $IF_{\lambda}$ and a PMT.

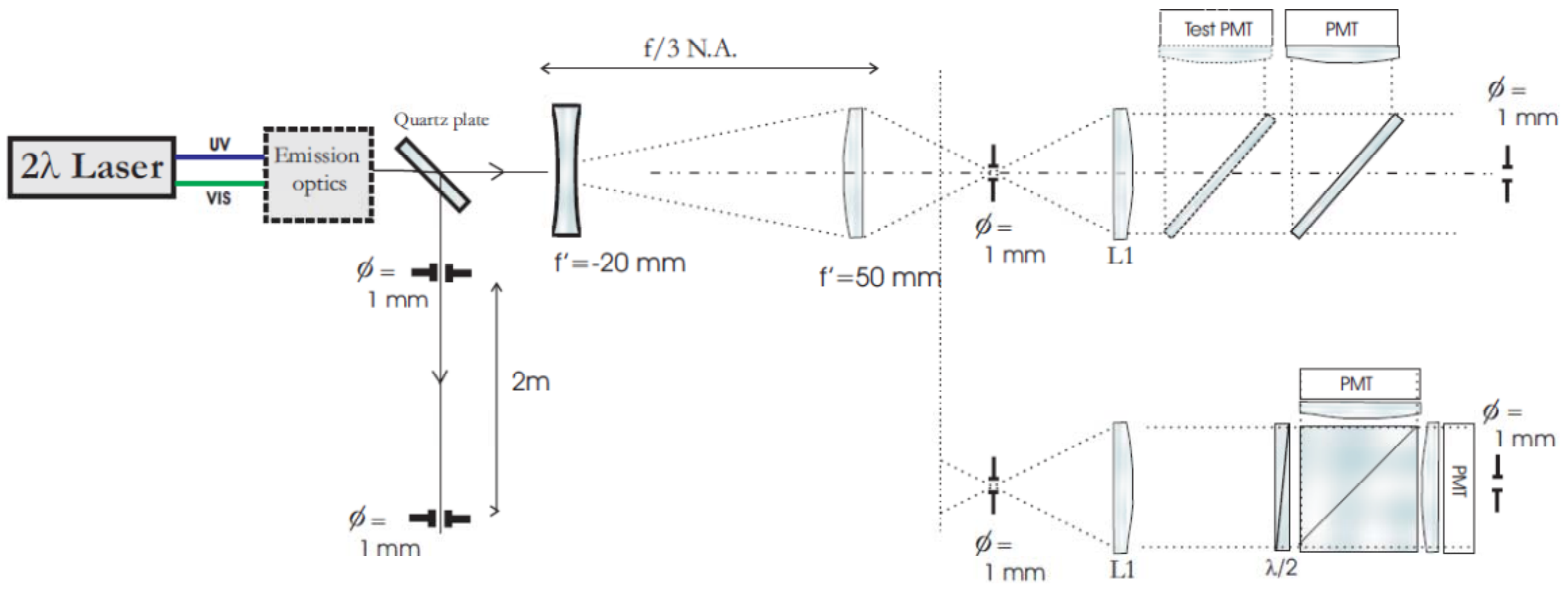

**Fig.5** Laboratory test-bench with numerical aperture f/3 to simulate backscattered photons from the atmosphere. The top scheme is used for measuring the $DB_{\lambda}$ reflectivity while the bottom scheme enables the $T_p$, $R_s$-measurements of the PBC's polarization properties, using the λ/2 plate to control the incident polarization.

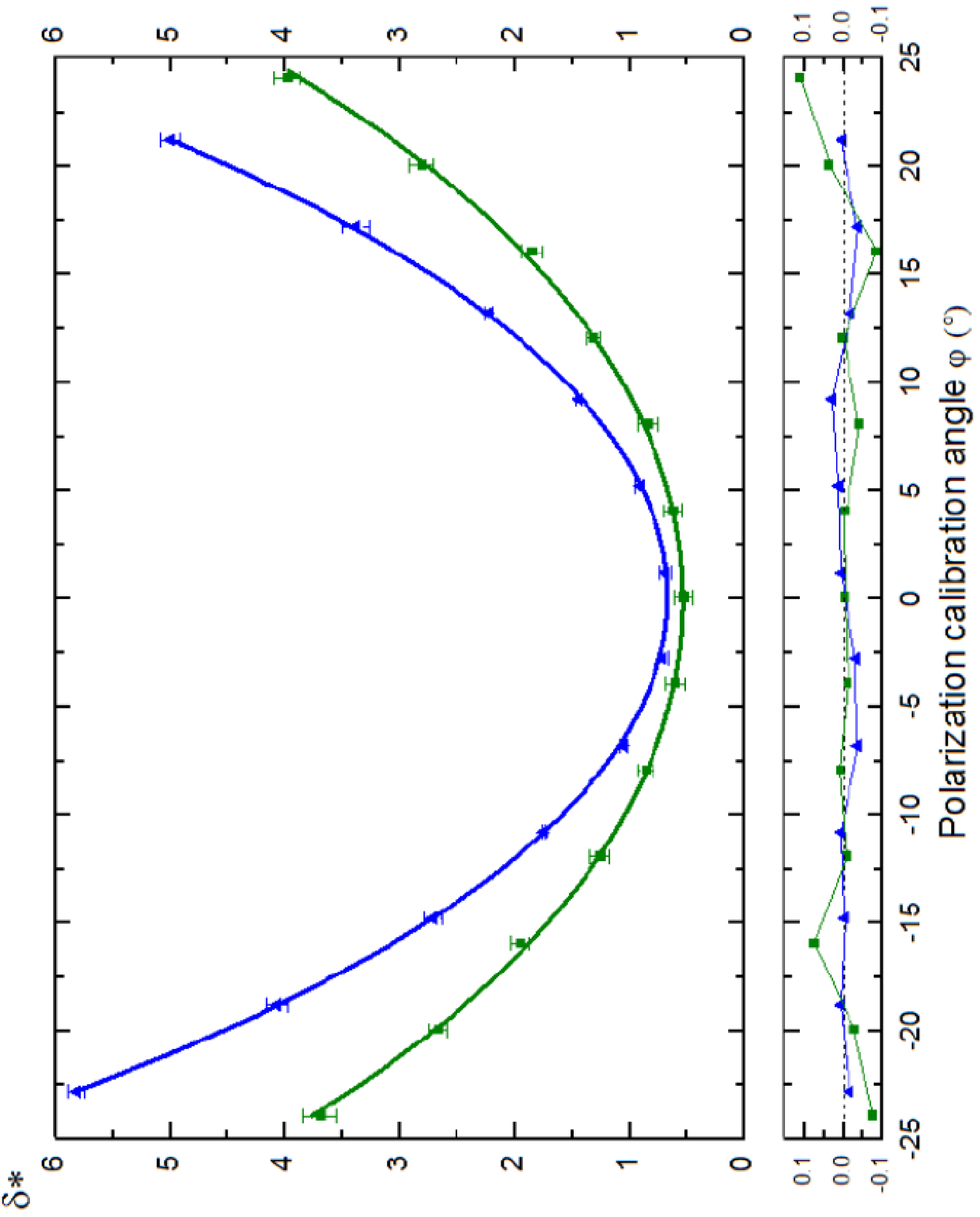

**Fig.6** Calibration curves obtained in the UV and the VIS with corresponding residue plot.

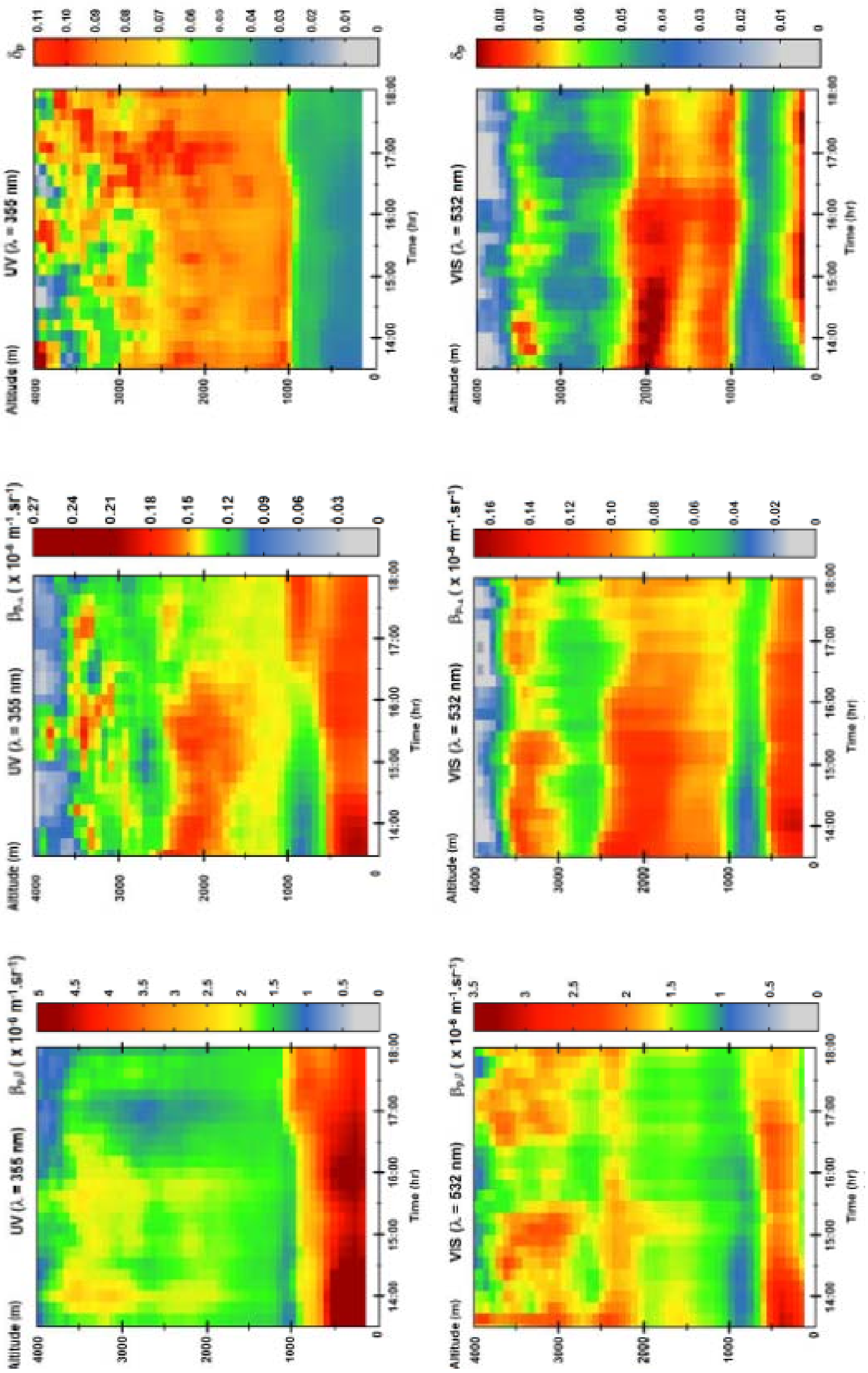

**Fig.7** Time-altitude maps of the parallel and perpendicular particle backscattering coefficient $\beta_{p,//}$ and $\beta_{p,\perp}$ and the corresponding particle depolarization ratio $\delta_p$, in the UV and the visible spectral range on October 18th 2011 at Lyon between 13h30 and 18h. The color scales have been adjusted on each map to enhance the achieved sensitivity in the UV and in the VIS. In between each 4000 laser shots-vertical profile, the laser has been shut down during 4 minutes.

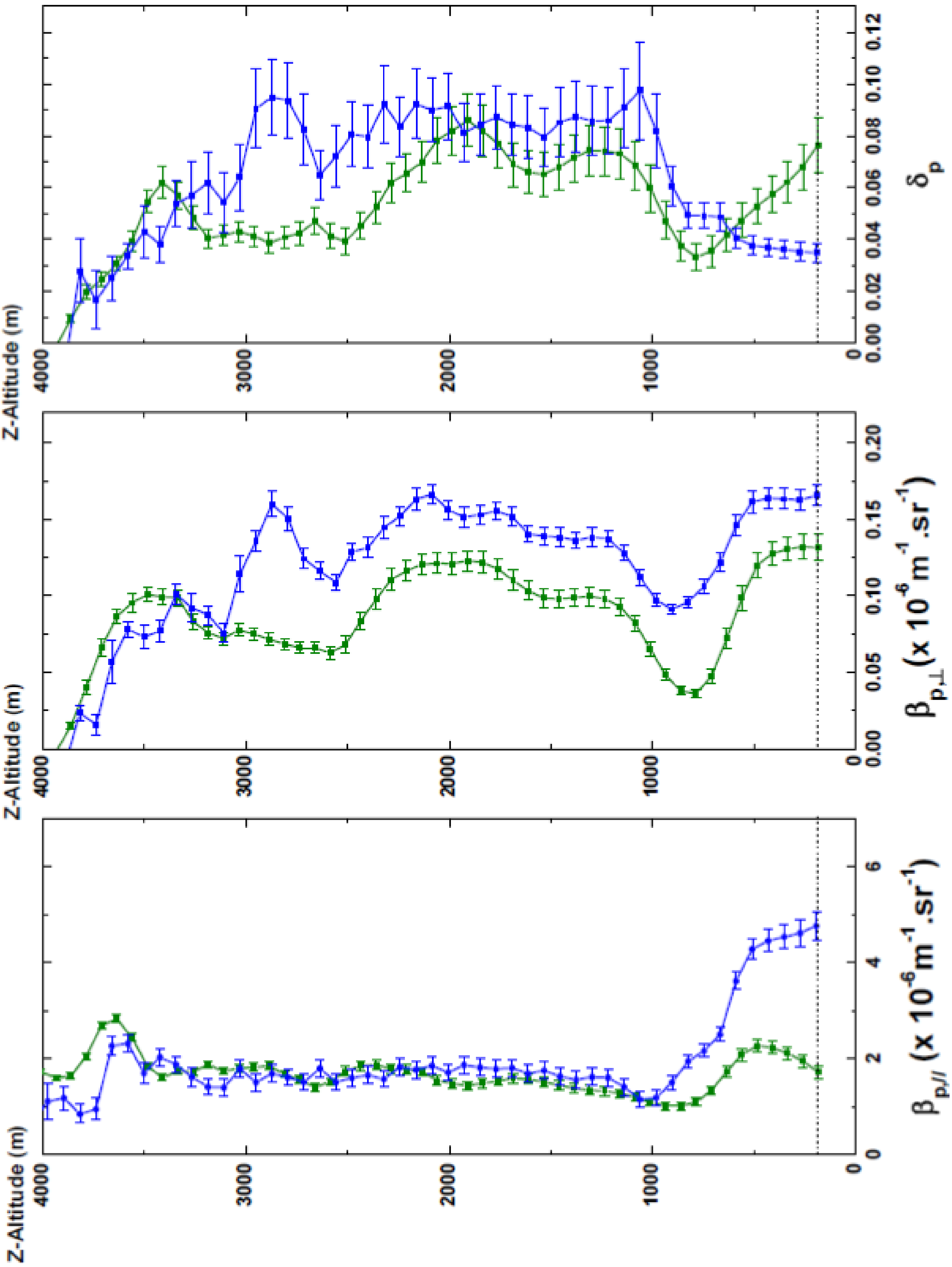

**Fig.8** Vertical profiles of $\beta_{p,//}$, $\beta_{p,\perp}$ and $\delta_p$ on October 18th 2011 at 14h45 at Lyon in the UV (blue) and the VIS (green). Error bars on $\beta_{p,//}$ are calculated by using the maximum and minimum values of $S_p$ in the Klett's algorithm. Error bars on $\beta_{p,\perp}$ are derived from the section 2-derived relation : $\beta_{p,\perp} = (R_{//}\delta - \delta_m) \times \beta_{m,//}$ while error bars on $\delta_p$ are calculated by applying equation (5).

| | | |
|---|---|---|
| **UV-VIS Lidar laser source** | | |
| Pulse energy | 10 mJ (UV), 20 mJ (VIS) | |
| Pulse duration | 10 ns | |
| Pulse repetition rate | 10 Hz | |
| Laser beam initial divergence | 1 mrad | |
| Laser initial polarization rate ($I_p/I_s$) | > 100:1 | |
| **Emitter optics** | | |
| Emission PBC (UPBS-λ-100) | $T_p/T_s > 250$ | $T_p/T_s > 500$ |
| Beam expanders (Bmx λ 2,5x) | × 2.5 | × 2.5 |
| Dichroic mirror (SWP-45-$R_s$532-$T_s$355-PW) | $T_s > 0.995$ | $R_s > 0.995$ |
| Emission mirror (NB1-K) | $R_s > 0.995$ | $R_p > 0.99$ |
| **Lidar receiver** | | |
| Primary mirror focal length | 600 mm | |
| Primary mirror diameter | 200 mm | |
| Secondary mirror diameter | 50 mm | |
| Pinhole diameter | 3 mm | |
| Field of view | 2.5 mrad | |

**Tab.1** Optical specifications of the emission laser, the emitter optics and the Lidar receiver.

| **UV-VIS polarization detector D** | | |
|---|---|---|
| Dichroic beamsplitter (355 nm) | $R_p$ = (72.3 ± 0.5) % | $R_s$ = (94.3 ± 0.5) % |
| Dichroic beamsplitters (532 nm) | $R_p$ = (99.9 ± 0.5) % | $R_s$ = (86.7 ± 0.5) % |
| PBC $_1$ (UPBS-UV-100) | $R_{s,1}$ = 1.000 | $T_{p,1}$ = 0.998 |
| PBC $_2$ (UPBS-UV-100) | $R_{s,2}$ = 0.980 | $T_{p,2}$ = 0.992 |
| PBC $_3$ (UPBS-UV-100) | $R_{s,3}$ = 0.998 | $T_{p,3}$ = 0.996 |
| $IF_\lambda$ center wavelength | 354.94 nm (UV) | 532.14 nm (VIS) |
| $IF_\lambda$ filter bandwidth | 0.35 nm (UV) | 0.52 nm (VIS) |

**Tab.2** Optical specifications of the UV-VIS polarization detector D.